\documentclass[aps,twocolumn,longbibliography,superscriptaddress]{revtex4-1}

\usepackage{amssymb}

\usepackage{color}
\usepackage{graphicx}

\newcommand{\textmark}[1]{\textcolor{black}{{#1}}}

\begin{document}

\begin{titlepage}



\title{Crossover in dynamics in the Kob-Andersen binary mixture glass-forming liquid}


\author{Pallabi Das}

\author{Srikanth Sastry}
\affiliation{Theoretical Sciences Unit, Jawaharlal Nehru Centre for Advanced Scientific Research,Jakkur Campus, Bengaluru, 560064,India}

\begin{abstract}
%

Glass-forming liquids are broadly classified as being fragile or strong, depending on the deviation from Arrhenius behavior of their relaxation times. A fragile to strong crossover is observed or inferred in liquids like water and silica, and more recently also in metallic glasses and phase change alloys, leading to the expectation that such a crossover is more widely realised among glass formers. We investigate computationally the well-studied Kob-Andersen model, accessing temperatures well below the mode coupling temperature $T_{MCT}$. We find that relaxation times exhibit a crossover in dynamics around $T_{MCT}$, and discuss whether it bears characteristics of the fragile to strong crossover. Several aspects of dynamical heterogeneity exhibit behavior mirroring the dynamical crossover, whereas thermodynamic quantities do not.  In particular, the Adam-Gibbs relation describing the relation between relaxation times and configurational entropy continues to hold below the dynamical crossover, when anharmonic corrections to the vibrational entropy are included.
\end{abstract}




\maketitle

\end{titlepage}


\section{Introduction}
\label{sec:intro}
The concept of {\it fragility} was introduced by Angell \cite{angell_fragility,angell1995formation,martinez2001thermodynamic} as a way of organizing the diversity of the remarkable slow down of dynamics in glass forming liquids as the glass transition is approached. Glass formers such as silica near the glass transition display an Arrhenius variation of viscosity, whereas other commonly investigated glass formers exhibit super-Arrhenius dependence of viscosity and relaxation times on temperature, to various degrees. Glass formers exhibiting Arrhenius temperature dependence are termed {\it strong} and those exhibiting super-Arrhenius temperature dependence as termed {\it fragile}. {\it Fragility} quantifies the degree of deviation from Arrhenius behavior, and has been investigated extensively \cite{fragility_book}. 

In attempting to rationalise experimental results close to the glass transition of water with those obtained at considerably higher temperatures in supercooled water, Angell\cite{angell_fragile_to_strong} proposed the possibility of a crossover from fragile behaviour at the higher temperatures to strong behaviour close to the glass transition. In addition to water \cite{ito1999thermodynamic,STARR200351,Shi9444,lupi2021dynamical}, such a fragile to strong transition has been investigated in computer simulations of silica  \cite{barrat1997strong,horbachsilica,saika2001fragile,heuersilica}, and silicon \cite{sastry2003liquid,vasisht2009study,jakseCp,vasisht2013phase}. In these liquids, all of which exhibit several well known anomalous properties arising from low density, open tetrahedral order that becomes dominant at low temperatures. The fragile to strong crossover is associated with a change in liquid structure towards more tetrahedral local geometries upon lowering temperature, a possible liquid-liquid transition \textmark{\cite{llptbook,saika2001fragile}}, and with the crossover being characterised by the presence of a heat capacity maximum. 

In recent years, the fragile to strong crossover in dynamics has also been reported in a variety of other glass formers, including  phase change alloys\cite{wei2015phase,orava2015fragile} used in memory devices, and several metallic glasses \cite{zhang2010fragile,wei2013,alvarez2019revisiting,SUKHOMLINOV2019129,zhang2021dynamic}. In many of these instances, though not all, the fragile to strong crossover has been shown to be accompanied by a heat capacity maximum. Unlike the cases such as silica discussed above, however, there is no broadly applicable picture of structural change that may drive such a crossover. 

Interestingly, a well studied computational model glass former that is commonly described as a {\it fragile} glass former, the Kob-Andersen $80:20$ binary Lennard-Jones mixture (KA-BMLJ), also exhibits a crossover in dynamics \cite{ashwin2003low,coslovich2018dynamic} that is reminiscent of the transition seen in computer simulations of silica \cite{horbachsilica}. Such a crossover has also been recently investigated for the same model potential but for different compositions ($2:1$ and $3:1$ rather than $4:1$)\cite{ortlieb2021relaxation} and for a soft sphere mixture\cite{flenner2013}. In the KA-BMLJ model, below the mode coupling temperature $T_{MCT}$, the relaxation times exhibit a crossover from non-Arrhenius to Arrhenius temperature dependence. We perform molecular dynamics simulations which are roughly an order of magnitude longer than those reported in \cite{coslovich2018dynamic}, and characterise the dynamics in detail, including several aspects of dynamical heterogeneity. We perform an analysis of the structure of clusters of mobile particles, to understand the changes in the geometry of such clusters across the dynamical crossover. We investigate the relation between the dynamics to thermodynamic changes. Specifically, we consider the behaviour of the heat capacity. Although the presence of a heat capacity maximum was reported in an earlier study \cite{flenner2006}, the recent work reported in \cite{coslovich2018dynamic} does not find evidence for such a maximum, a result which we confirm to hold to lower temperatures. We also investigate the validity of the Adam-Gibbs relation, found to be valid in computer simulations of the KA-BMLJ \cite{Sastry2001,karmakar2009growing,sengupta2012adam} as well as other systems\cite{scala2000water,saika2001fragile,saika2001fragile,starr2013relationship}. However, 
recent work\cite{ozawa2019does} concludes that a generalised form of the Adam-Gibbs relation is required to describe the relaxation times at temperatures significantly lower than previously investigated, as also observed for two dimensions in \cite{sengupta2012adam}. We find deviations from the Adam-Gibbs prediction below the dynamical crossover temperature, when the configurational entropy is evaluated with a harmonic approximation to the vibrational entropy (as has been done in the past for the studied system \cite{Sastry2001,karmakar2009growing,sengupta2012adam}). However, when anharmonic correction to the vibrational entropy estimates are included, the relaxation times are found to obey the Adam-Gibbs relation across all the temperatures investigated. 

We first describe the model studied and the simulation details, followed by a description of the simulation results. We conclude with a discussion of the significance and implication of these results.

\section{Model and Simulation Details}
\label{mm}
We study the Kob-Andersen (KA) $80:20$ binary mixture \cite{KAref}, with the interaction potential between particles given by 

\begin{eqnarray}
U_{\alpha\beta}(r) &=& 4\epsilon_{\alpha\beta}\left[\left(\frac{\sigma_{\alpha\beta}}{r}\right)^{12}-\left(\frac{\sigma_{\alpha\beta}}{r}\right)^{6}\right]  \nonumber \\
                     && \quad+ 4\epsilon_{\alpha\beta}\left[c_{0}+c_{2}\left(\frac{r}{\sigma_{\alpha\beta}}\right)^{2}\right] , r_{\alpha\beta} \leq r_{c~\alpha\beta} \nonumber \\
                     &=& 0, \hspace{3.0cm} r_{\alpha\beta} > r_{c~\alpha\beta}.
 \end{eqnarray}
 Here, $c_0$ and $c_2$ are chosen such that the potential and force between particles go to zero continuously at the cutoff distance $r_{c~\alpha\beta}$ \textmark{$(=2.5\sigma_{\alpha\beta})$}. The indices $(\alpha, \beta)$ represent particle type ($A$ or $B$) in the binary mixture. We report results in reduced units, with units of length, energy and time scales being $\sigma_{AA}$, $\epsilon_{AA}$  and  $\sqrt {{\sigma^2_{AA}m_{AA}}/{\epsilon_{AA}}} $, respectively. The model parameters are 
  $\epsilon_{AB}/\epsilon_{AA}  = \epsilon_{BA}/\epsilon_{AA} = 1.5$, $\epsilon_{BB}/$ $\epsilon_{AA}  = 0.5 $, and
$\sigma_{AB}/\sigma_{AA} = \sigma_{BA}/\sigma_{AA} = 0.8 $, $\sigma_{BB}/\sigma_{AA} = 0.88 $. Energy values reported are energies per particle. Constant volume, temperature (NVT) simulations have been performed for the system size $N = 4000$ \textmark{using the Nos\'{e}--Hoover thermostat}. The initial configurations for the simulation below $T = 0.466$ are prepared by quenching equilibrated configurations from $T = 0.466$ and for the simulations above $T = 0.466$  configurations are obtained by quenching equilibrated configurations from $T = 1.0$ to respective target temperatures. We have performed equilibrium simulation for a range of temperatures, from $T = 0.9$ to $T = 0.365$. The number density ($\rho = N/V$, where $V$ is the volume) has been kept constant at $1.2$. At each temperature,  $16 - 24$ independent trajectories have been studied. The time step below and above the mode coupling temperature $T_{MCT} = 0.435$ are respectively $dt = 0.01$ and $dt = 0.005$.  \textmark{We note, however, that the use of the larger time step at low temperatures leads to a shift in the per particle energy of $\approx 5 \times 10^{-3}$. We perform additional runs with the smaller time step of $0.005$ to rectify this shift in the energy values employed.}
At the lowest temperatures, run lengths extend up to $2 \times 10^{10}$ integration time steps or a time duration of $2\times 10^{8}$. All simulations have been performed using LAMMPS \cite{plimpton1995fast}. The relevant quantities are reported for the $A$ type of particles unless otherwise mentioned. \textmark{The system is prone to crystallization at low temperatures. In the range of temperatures from $T_{MCT}$ to the lowest simulated temperature, the percentage of runs that crystallize increases from $5\%$ to $80\%$}. The crystallizing samples have been identified using standard methods employing bond orientational order parameters\cite{rein1996numerical}, as described in \cite{das2018annealing,coslovich2018dynamic,PhysRevXCryst2019} and discarded from the analysis.

\section{Results}

We present results concerning the structural relaxation times, and several measures of dynamical heterogeneity, which display a crossover around the mode coupling temperature. We describe next the results concerning the thermodynamic changes with temperature and their relationship with the observed dynamical crossover. Details of the definitions of the quantities investigated and supporting data are provided in appendices in order to streamline the presentation of the main results. 


\subsection{Structural relaxation times}

\begin{figure}[t]
\centering
\includegraphics[scale=.45]{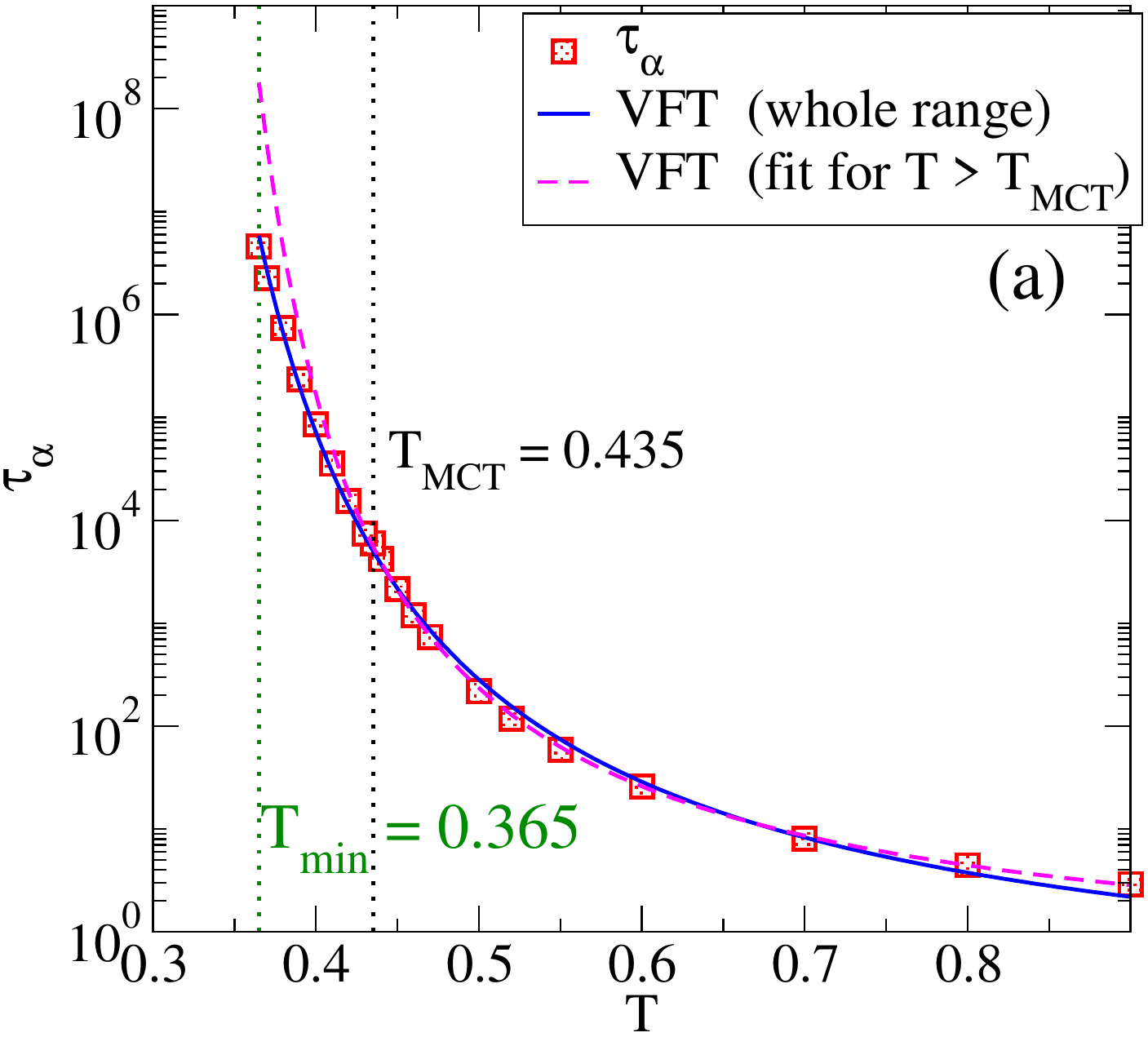}
\includegraphics[scale=.45]{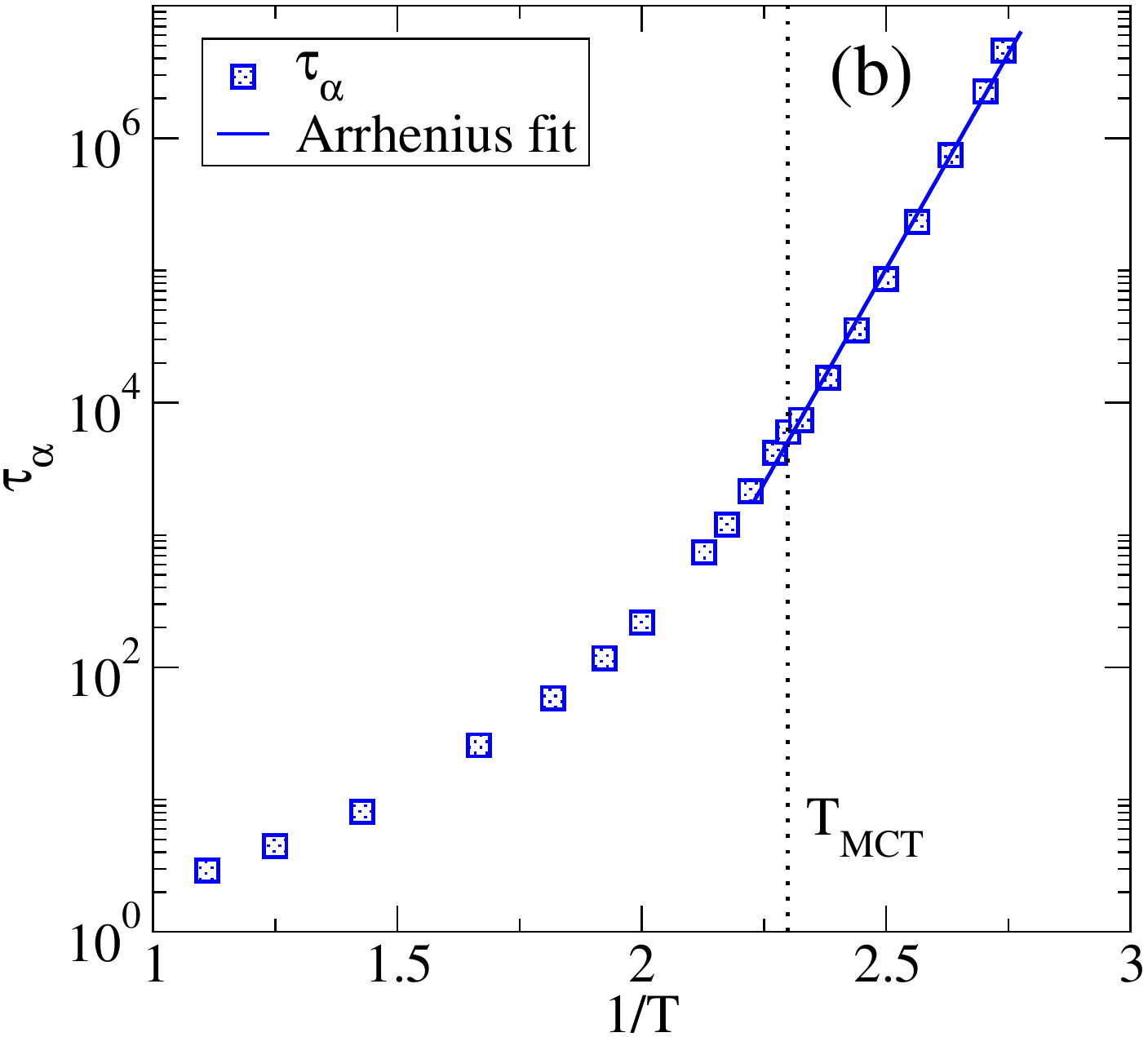}
\caption[Relaxation times]{(a) Relaxation times $\tau_{\alpha}$ extracted from the decay of the overlap $q(t)$ for temperatures across $T_{MCT}$ are shown, along with VFT fits to data for $T > T_{MCT}$ (red dashed line) and the full range of $T$ values (blue solid line). The fit to high temperature data overestimates relaxation times at lower temperatures, whereas the fit to the entire range shows deviations from the data points at the higher temperatures. 
(b) Relaxation times $\tau_{\alpha}$  from the decay of $q(t)$ {\it vs.} $1/T$. \textmark{The fit line through the data for $T < T_{MCT}$ demonstrates that at low temperatures, relaxation times can be better approximated by an Arrhenius temperature dependence.} }
\label{vftarrh}
\end{figure}

Structural relaxation times are computed by considering the overlap function $q(t)$, defined in \ref{appendixA} and the self intermediate scattering function $F_s(k,t)$, described in \ref{appendixB}. 
These time correlation functions (shown in \ref{appendixA} and \ref{appendixB}) are fitted with a four parameter functional form, which is expressed for $q(t)$ as 
\begin{equation} \label{qkww}
q(t) = (1-f_c)exp(-(t/\tau_s))^n+f_cexp(-t/\tau_{\alpha})^{\beta_{kww}}
\end{equation} 

where $f_c$ is the non-ergodicity parameter, $\tau_{\alpha}$ is the structural relaxation time, $\beta_{kww}$ is the Kohlrausch-Williams-Watts stretching exponent, and $\tau_{s}$ is a relaxation time that describes the short time decay of the correlation functions. The exponent $n$ describing the short time decay, based on results in  \cite{sengupta2013breakdown}, is chosen to be $n=2$. 
The $\tau_{\alpha}$ extracted from the fitted form for $q(t)$ are plotted as the function of the temperature (see Fig. \ref{vftarrh}). 

We next fit the structural relaxation time $\tau_{\alpha}$ with the the Vogel Fulcher Tamman relation (VFT) expression 

\begin{equation} 
 \tau = \tau_0 \exp \left [ \frac{1}{(K_{VFT}({T}/{T_{VFT}}-1))}\right],
 \end{equation} 
 that is often employed to describe relaxation times in glass formers. Previous work has estimated  the mode coupling temperature $T_{MCT}$  to be $T=0.435$ from power law fits of relaxation times for higher temperatures  ($\tau=\tau_0(T-T_{MCT})^{-\gamma}$) \cite{KAref}. In Fig. \ref{vftarrh} (a), we show VFT fits to the relaxation times, by considering data only for $T > T_{MCT}$ \textmark{$(\tau_0 = 0.3101$, $K_{VFT} = 0.2243$, $T_{VFT} = 0.2989$)}, as well as the entire range available. The VFT fits to the high temperature data clearly overestimate the relaxation times for $T < T_{MCT}$. On the other hand, for the VFT fit to the full range \textmark{($\tau_0 = 0.1175$, $K_{VFT} = 0.1383$, $T_{VFT} = 0.2592$.)}, the VFT form does not provide a good description of the data at higher temperatures. Fig. \ref{vftarrh} (b) shows the same results in an Arrhenius plot, which shows that at temperatures below $T_{MCT}$, \textmark{the slower increase of relaxation times can be better approximated by an Arrhenius form.} \textmark{The value of the activation energy barrier from the fit is $\sim 15$, which agrees well with the the previously reported value in \cite{coslovich2018dynamic} at which the activation energy shows a possible saturation.} Such a crossover from super-Arrhenius to Arrhenius behavior has indeed previously been observed \cite{ashwin2003low,coslovich2018dynamic,ortlieb2021relaxation}. Our results extend the range of temperatures explored.

 


\subsection{Dynamical heterogeneity}

\begin{figure*}[t]
\centering
\includegraphics[scale=.45]{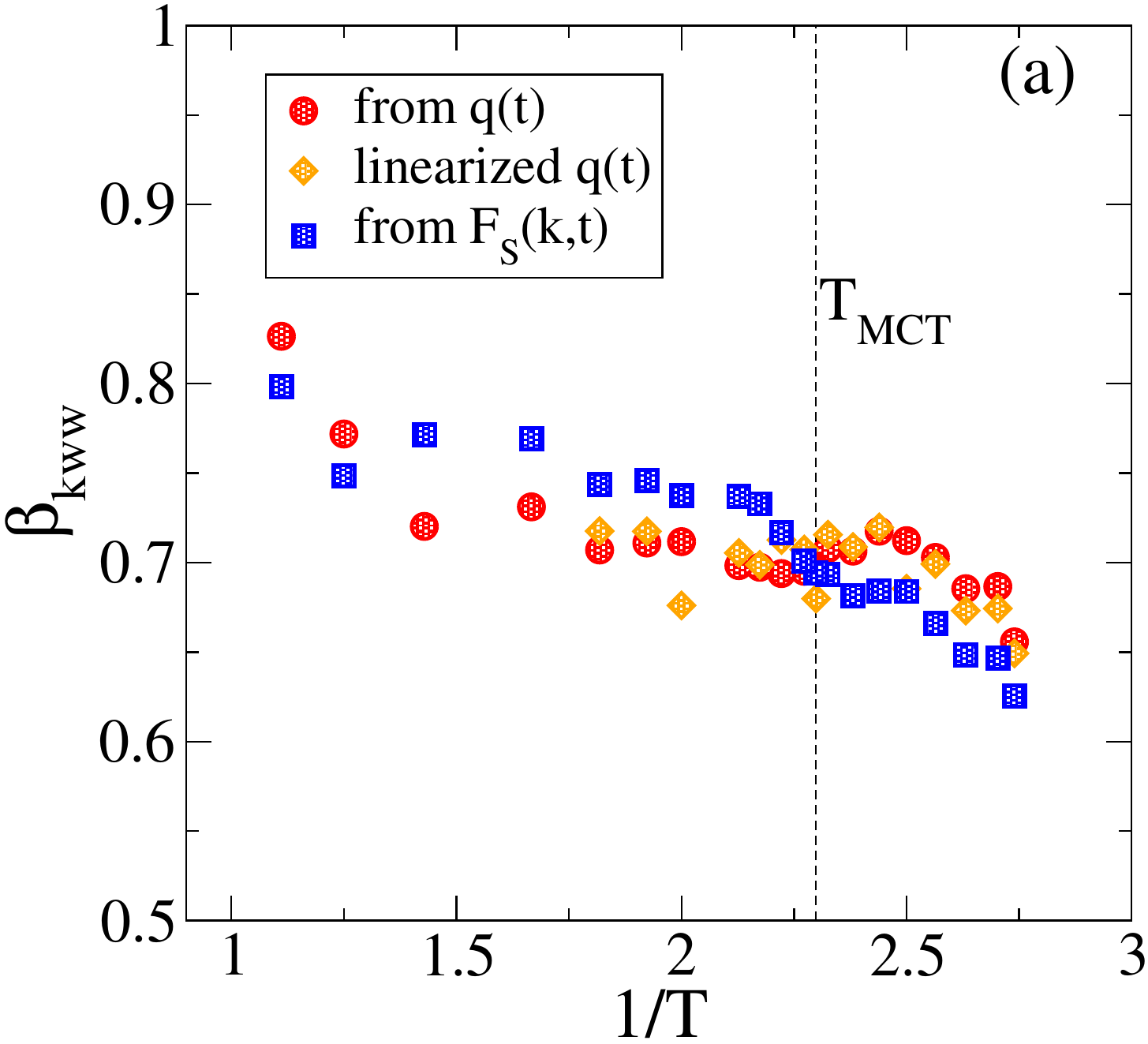}
        \includegraphics[scale=.45]{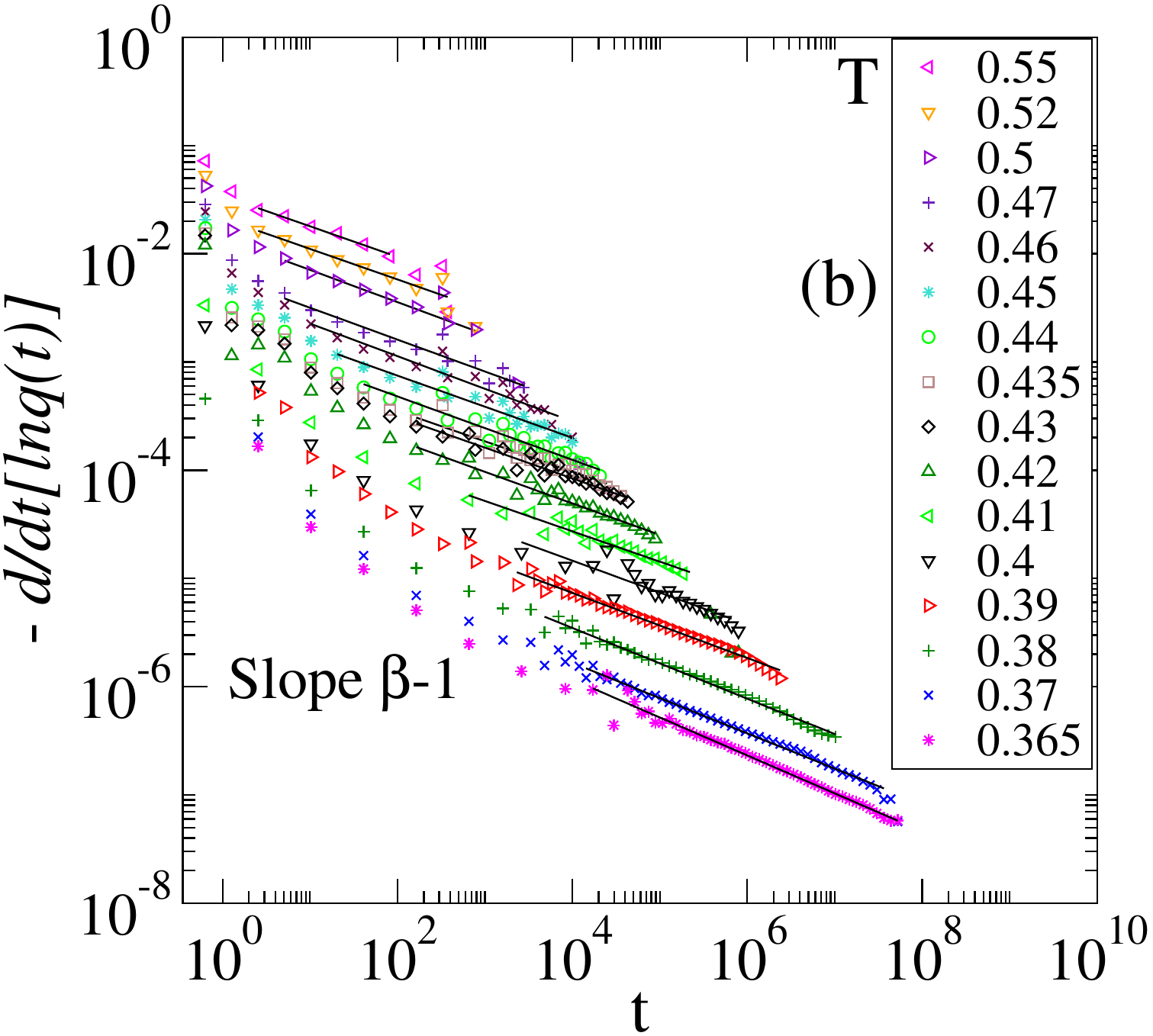}
\caption{(a) The stretching exponent $\beta_{kww}$ is shown here as a function of the temperature, obtained from stretched exponential fits of the overlap function $q(t)$ and $F_s(k,t)$, as well as by transforming the $q(t)$  to obtain $\beta_{kww}$ from the resulting linearisation of the data. (b) The transformation of $q(t)$ used to estimate $\beta_{kww}$ from the linearised data. The smallest $q(t)$ value used is $q(t) = 3 \times 10^{-3}$. 
}
\label{beta}
\end{figure*}

We next discuss three quantities that are associated with dynamical heterogeneity, namely the stretching exponent in the stretched exponential fits to the relaxation functions, $\beta_{KWW}$, the non-Gaussian parameter $\alpha_2$, and the dynamical susceptibility $\chi_4$.

\subsubsection{Stretching exponent ($\beta_{kww}$)} 

The Kohlrausch-William-Watts (KWW) stretched exponential form exhibited by correlation functions ($q(t)$ and $F_s(k,t)$ in our study) have been investigated as a manifestation of heterogeneous dynamics \cite{EdigerReview,sengupta2013breakdown}. We extract the exponent $\beta_{kww}$ from the fits as expressed in Eq. \ref{qkww}, from $q(t)$ as well as $F_s(k,t)$ at $k = 7.25$, as shown in Fig. \ref{beta} (a).
We also estimate $\beta_{kww}$ by linearizing the long time behavior of $q(t)$, which can be written as $q(t) = f_c exp(-(t/\tau)^{\beta})$. From this, we can write (along the lines in \cite{ogielsky}) 

\begin{equation}
\label{eqnog}
\ln[-\frac{d}{dt}[\ln~q(t)]] = \ln(\frac{\beta}{\tau^{\beta}})+(\beta -1)ln ~t
\end{equation}


The stretching exponent $\beta_{kww}$ can be obtained from this linearised form as the slope of $\ln[-\frac{d}{dt}[\ln~q(t)]]$ $vs$ $\ln(t)$. The transformed values $ln[-\frac{d}{dt}[\ln~q(t)]]$  $vs$ $\ln(t)$ are shown in Fig. \ref{beta} (b) and the extracted $\beta_{kww}$ values are shown in Fig. \ref{beta} (a). The obtained values are noisy but, after an initial decrease from $1$ as temperature is lowered, $\beta_{kww}$ values show a mild decrease as temperature is decreased, with most temperatures shown displaying an exponent value of $\sim 0.7$. With the estimates available, it is difficult to draw any conclusions about a possible change in behavior across $T_{MCT}$. The exponent values from $F_s(k,t)$ at other $k$ values are shown in \ref{appendixB}.



\subsubsection{Non-Gaussian parameter ($\alpha_2(t)$)}

The non-Gaussian parameter  $\alpha_2(t)$ is defined as 
\begin{equation}
\alpha_2(t)= \frac{3<r^4(t)>}{5<r^2(t)>^2}-1
\end{equation}
and is a measure of how non-Gaussian the distribution of single particle displacements is. In the case of normal diffusive dynamics the displacement distribution is Gaussian and $\alpha_2$ is zero. In glass forming liquids, below the onset temperature \cite{sastry1998signatures,sastry2000onset}, the parameter goes through a maximum at a characteristic time $t^{*}$ and is zero in the short and the long time limits\cite{kob1997dynamical,sengupta2013breakdown,starr2013relationship}. The peak value is taken as a measure of the degree of heterogeneity, which increases with the decrease of the temperature. The time at which the peak occurs, $t^{*}$, is considerably shorter than the alpha relaxation time $\tau_{\alpha}$. 
Analysis presented in \cite{starr2013relationship} showed that the heterogeneity reflected in the behavior of $\alpha_2(t)$ is associated with clusters of spatially correlated {\it mobile} particles, and that the time scale at which such heterogeneity is maximum is closely related to the diffusive time scale $\left(D/T\right)^{-1}$. We will return to these considerations below. 

\begin{figure*}[t]
\centering
\includegraphics[scale=.4]{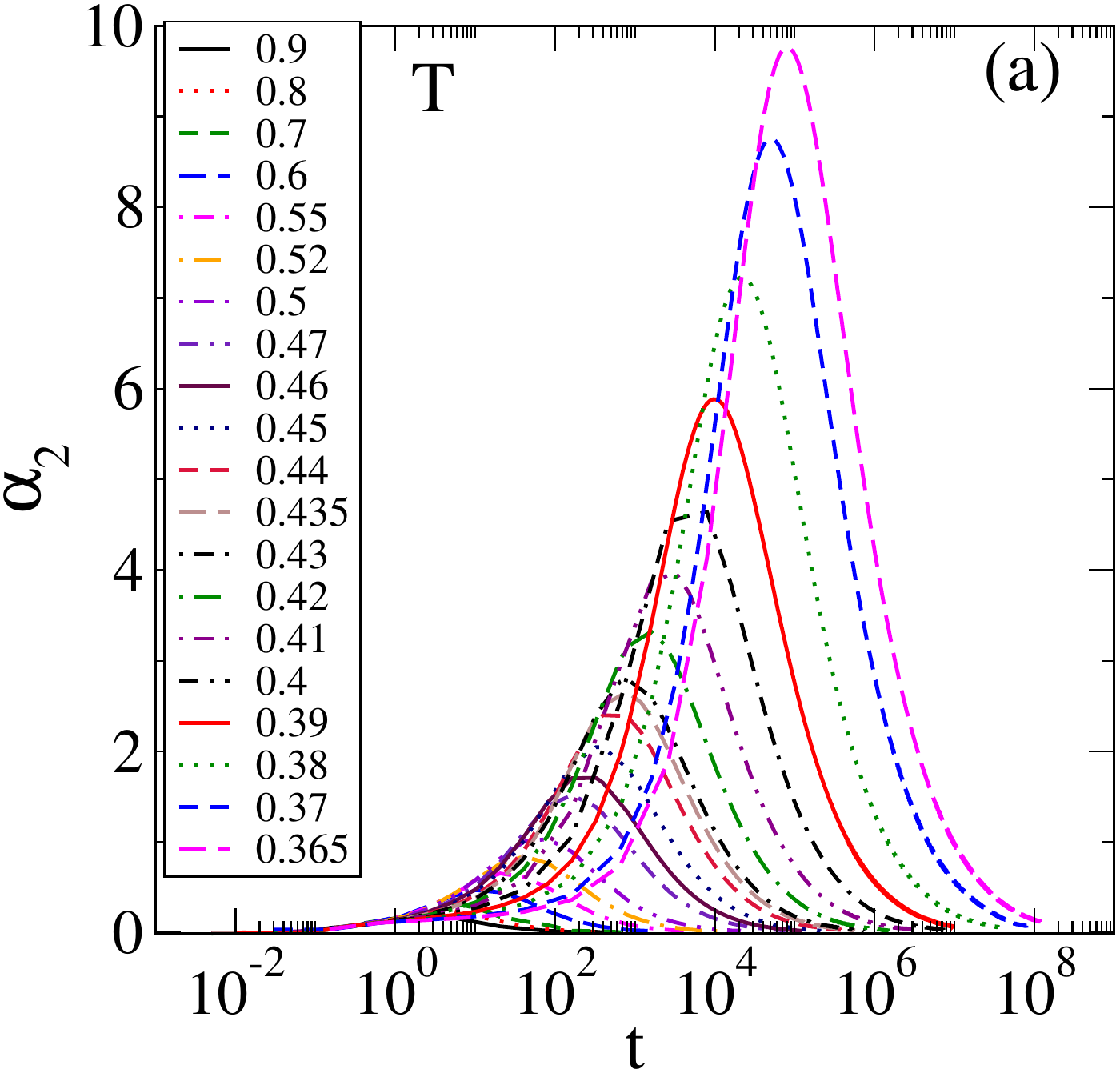}
\includegraphics[scale=.4]{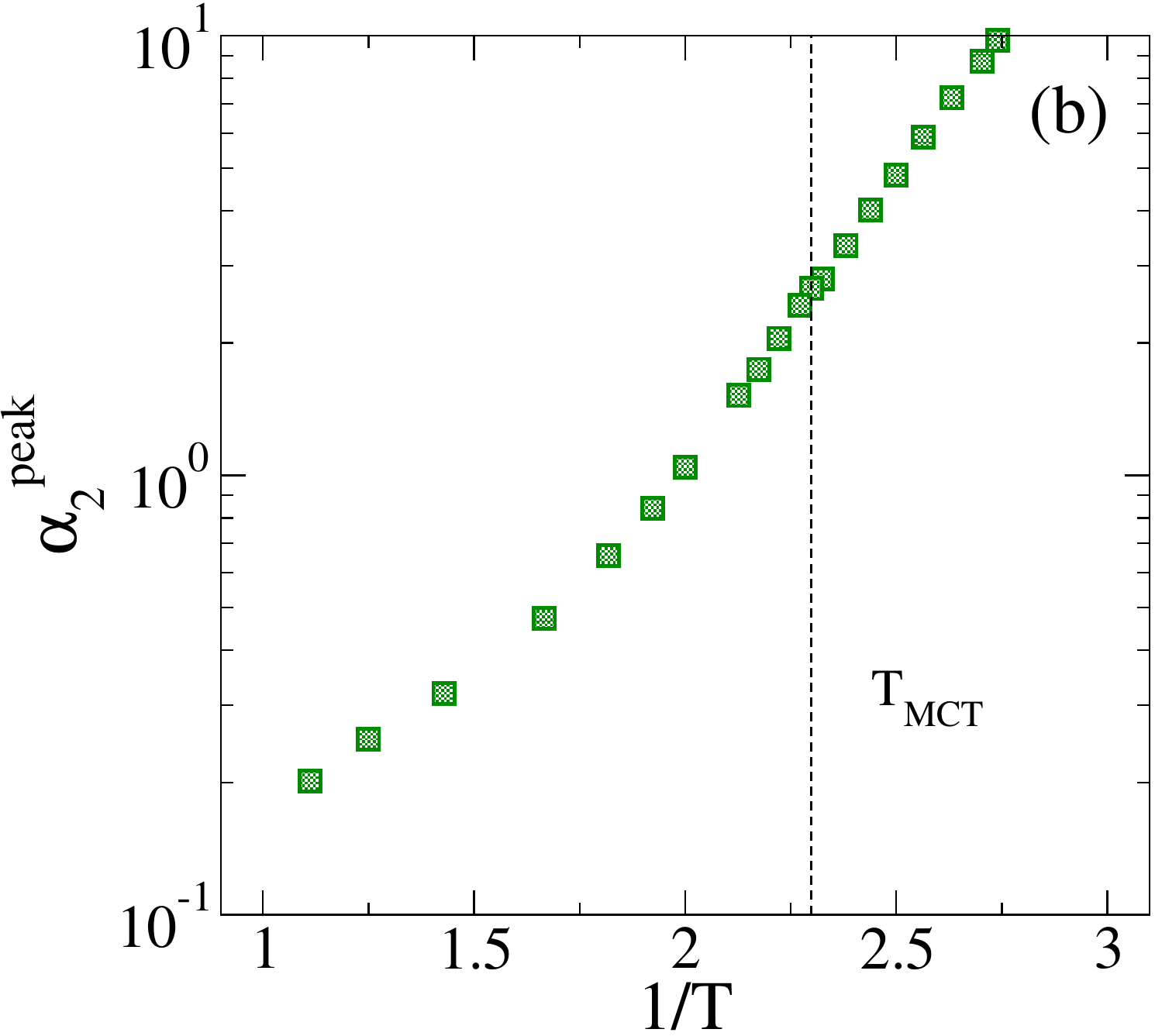}
\caption{(a) The non-Gaussian parameter $\alpha_2(t)$ as a function of time. The temperatures are indicated in the legends. (b) The peak value of the non-Gaussian parameter $\alpha_2^{peak}$, increases monotonically upon lowering temperature, but displays a change in the manner of increase around $T_{MCT}$. 
}

\label{alpha2}
\end{figure*}

Fig. \ref{alpha2} (a) shows the $\alpha_2(t)$ values for the range of temperatures studied, indicating that both the peak value $\alpha_2^{peak}$, and the time at which it occurs, $t^{*}$, increases upon lowering temperature.  We will discuss the behavior of $t^{*}$ further below when we compare different time scales emerging from our study. Fig. \ref{alpha2} (b) shows the temperature dependence of the peak value $\alpha_2^{peak}$, which displays a change in the temperature dependence below $T_{MCT}$, with values at lower temperatures falling below values one may expect from an extrapolation of the high temperature behavior.

\subsubsection{Dynamical susceptibility ($\chi_4$)}

\begin{figure*}[]
\centering
\includegraphics[scale=.4]{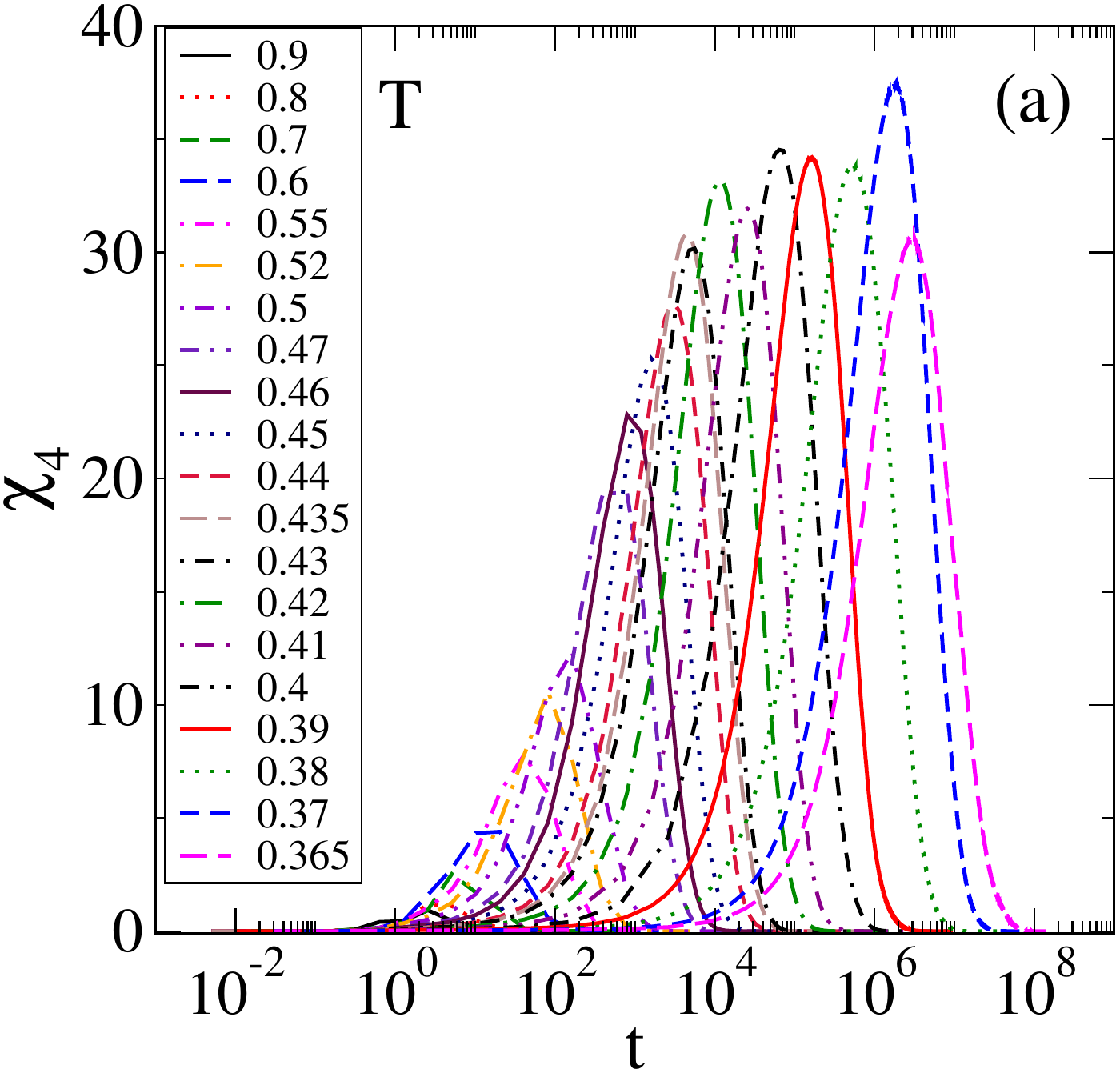}
\includegraphics[scale=.4]{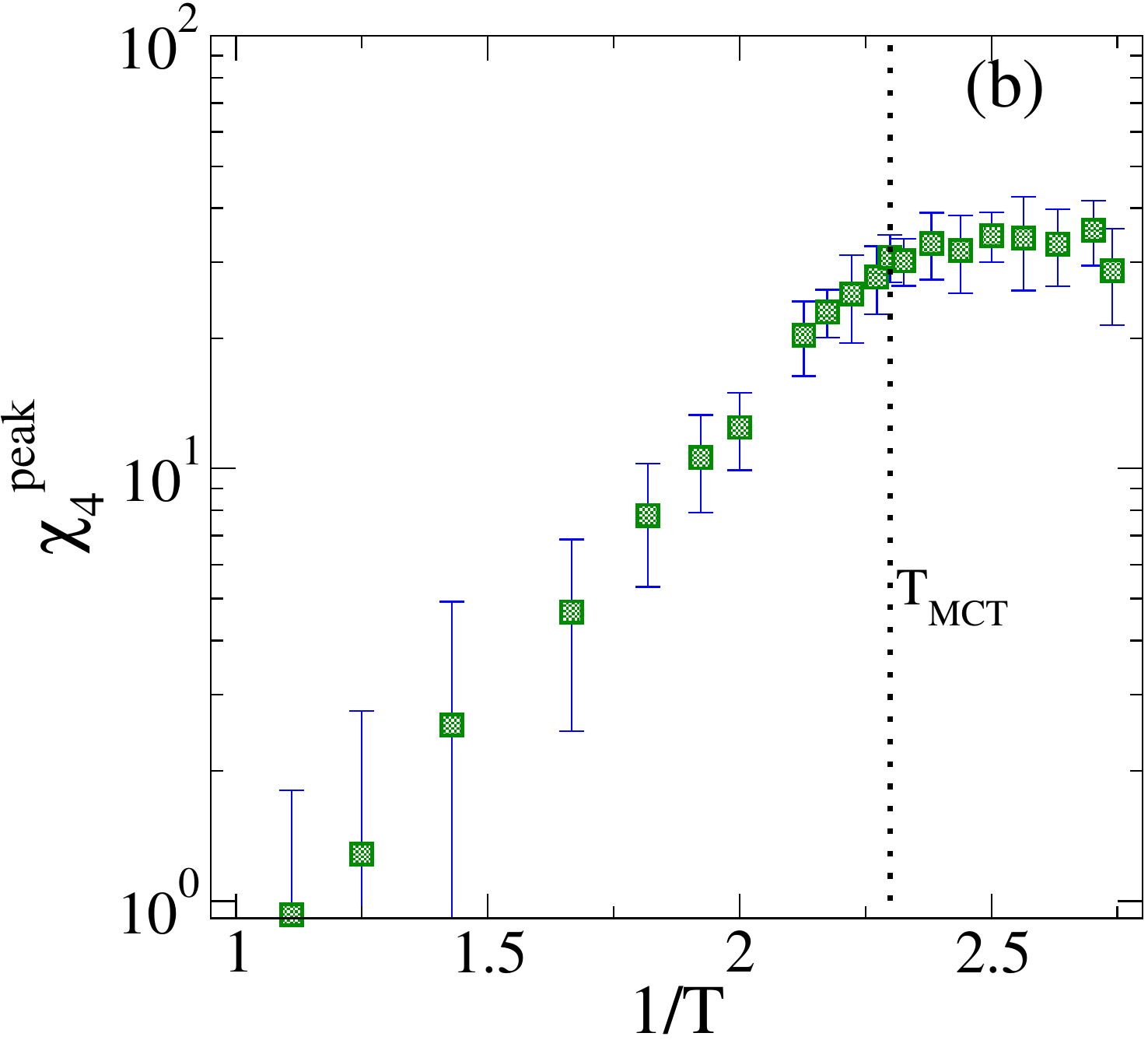}
\caption{(a) The dynamical susceptibility $\chi_4$ as a function of time. The temperatures are indicated in the legends. (b) The peak value of $\chi_4$ shows a saturation at the lower temperatures. 
}
\label{chi4}
\end{figure*}

The dynamical susceptibility $\chi_4(t)$ measures the fluctuations of the overlap function $q(t)$, and is defined (considering only $A$ particles) as 
\begin{equation}
\chi_4(t) = N_A[<q(t)^2>-<q(t)>^2].
\end{equation}
The peak of the $\chi_4(t)$\cite{glotzer2000time,karmakar2009growing} gives a measure of the amount of heterogeneity in the system. In Fig. \ref{chi4} (a), we show $\chi_4(t)$ for the different temperatures studied. Like $\alpha_2(t)$, $\chi_4(t)$ exhibits a peak at a characteristic time $\tau_4$. We compare $\tau_4$ with other characteristic time scales below. 
The peak height $\chi_4^{peak}$, shown in Fig. \ref{chi4} (b), clearly displays a crossover in behavior upon crossing $T_{MCT}$, becoming nearly constant, and possibly exhibiting a weak maximum. Our results are consistent with those obtained in \cite{coslovich2018dynamic}, and the change in behavior of $\chi_4(t)$ observed for the $2:1$ and $3:1$ KA-BMLJ model reported in \cite{ortlieb2021relaxation}. In recent work on a metallic glass  model \cite{zhang2021dynamic}, a clear peak in the  $\chi_4^{peak}$ values is observed, along with a specific heat maximum, which are described as manifestations of a fragile to strong crossover.  As described later and in \cite{coslovich2018dynamic}, no evidence of a specific heat maximum is found in the present model. Further, it is argued in \cite{coslovich2018dynamic} that the occurrence  of the crossover near $T_{MCT}$ may be accidental. Our results are not able to clarify the issue further, although the change in behavior in  $\chi_4^{peak}$ occurs quite convincingly when $T_{MCT}$ is crossed.



To summarise briefly the results so far, we see a change in the nature of dynamical heterogeneity when the temperature is decreased below $T_{MCT}$, most convincingly in the case of $\chi_4$, but also in the case of $\alpha_2(t)$. The stretching exponent $\beta_{kww}$  results we have are sufficiently noisy that we can not draw any conclusions of a crossover in behavior, although they do indicate that the dynamics becomes more heterogeneous as temperature decreases. 



 \subsection{Mobile particle clusters and strings}
 \begin{figure*}[t]
\centering
\includegraphics[scale=.35]{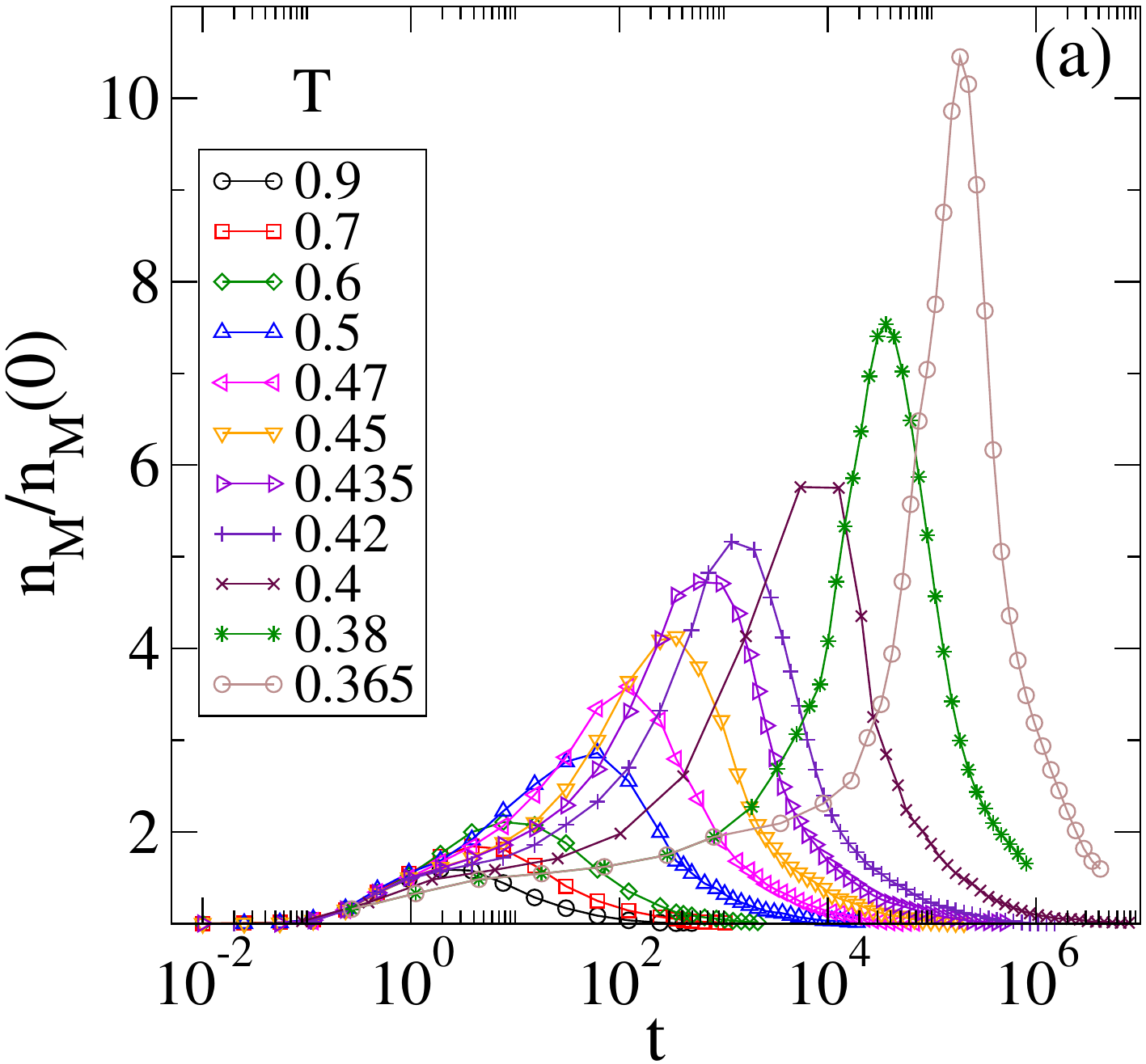}
\includegraphics[scale=.35]{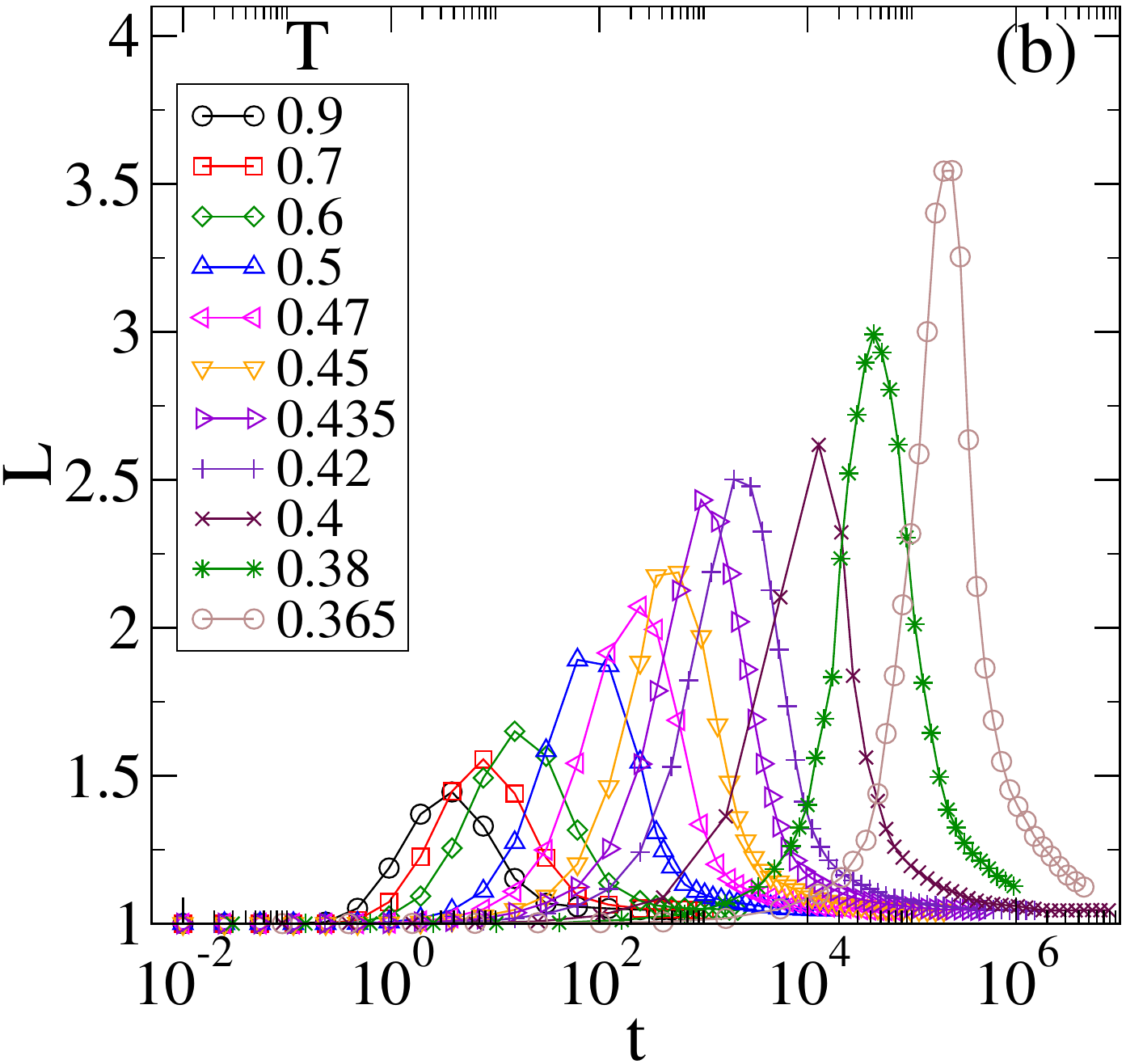}
\includegraphics[scale=.35]{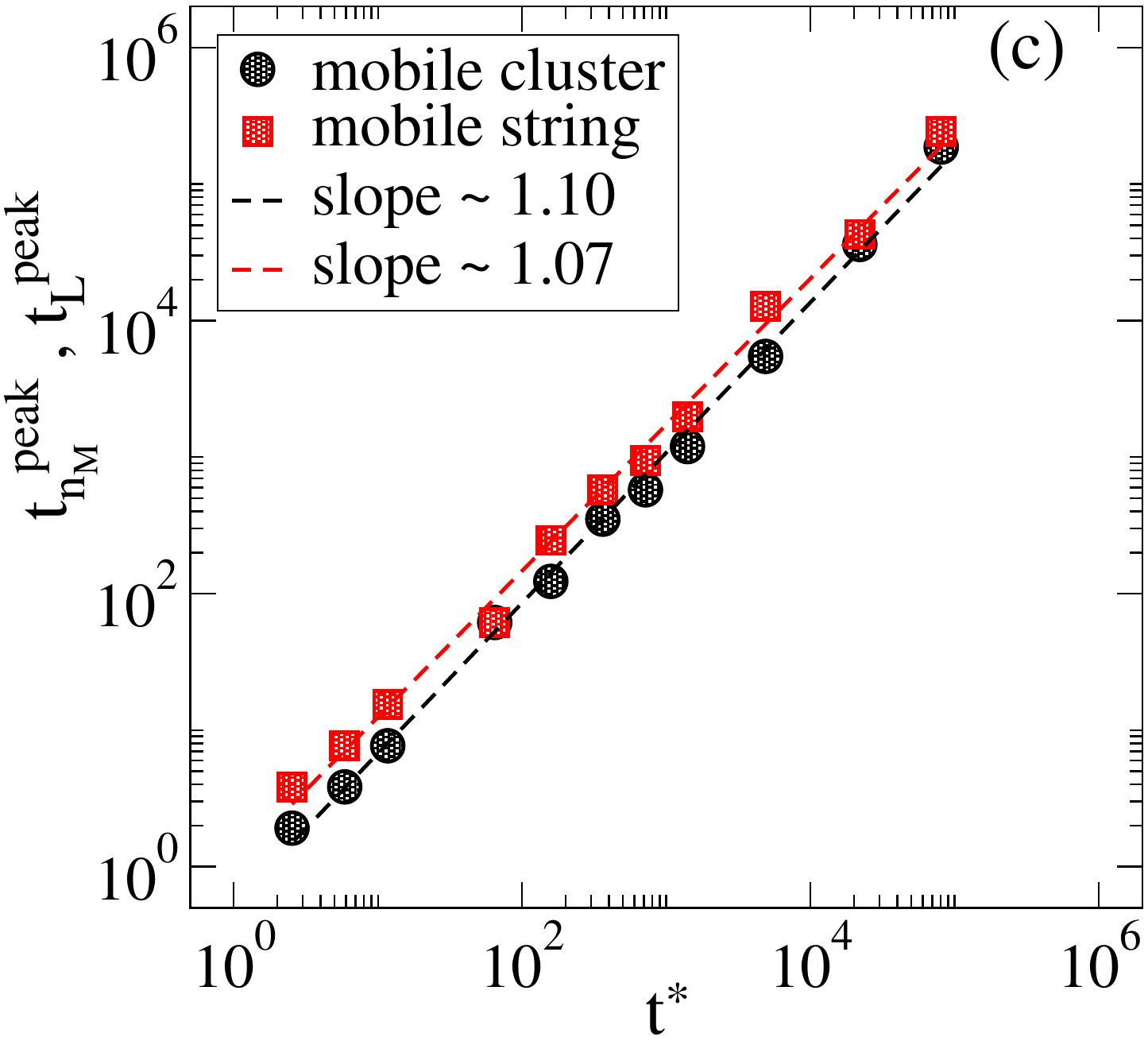}
\caption{The average size of mobile clusters and strings, shown in (a) and (b), exhibit maxima  at characteristic time scales $t_{n_M}^{peak}$ and $t_L^{peak}$ respectively. (b) The time scales  $t_{n_M}^{peak}$ and $t_L^{peak}$ plotted as functions of time $t^{*}$.
}
\label{mobstringtevolve}
\end{figure*}

We next consider the statistics and morphology of clusters of particles that move in a correlated fashion, which have been a subject of considerable study \cite{donati1998stringlike,PhysRevE.60.3107,glotzer2000time,appignanesi,stevenson2006shapes,karmakar2014growing,starr2013relationship,ortlieb2021relaxation,zhang2021dynamic}. In particular, the time dependent size of spatially correlated clusters of mobile and immobile particles, as well as string-like correlated mobile particles were investigated in \cite{starr2013relationship}, as also the morphologies of such clusters. It was found that the mean size of mobile particle clusters as well as strings displayed a non-monotonic time dependence, exhibiting a peak size at a time scale $t_{n_M}^{peak}$ and $t_L^{peak}$ respectively. These time scales were found to closely track the time $t^{*}$ at which $\alpha_2(t)$ exhibits a maximum, and in turn, the diffusion time scale $\left(D/T\right)^{-1}$. It was found further in \cite{starr2013relationship} that clusters of immobile particles exhibit a maximum mean size at a time that corresponds to the alpha relaxation time. 
In the present work, we do not investigate immobile particle clusters, but focus attention on clusters of mobile particles and strings. 

The large values of the non-Gaussian parameter for times $\sim t^{*}$ has been shown in several works \cite{kob1997dynamical,PhysRevE.60.3107} to correspond to the presence of a sub-population of particles that move much farther than the remaining particles. Such particles have further been shown to be spatially correlated. In these investigations, analysis of spatially correlated clusters was performed by considering the top $5 \%$ most mobile particles, which we also follow here (other works employ slightly different fractions; while the choice of the subset of mobile particles is thus arbitrary, qualitative behavior that emerges from such choices is not sensitively dependent on the choice). For any given time $t$, {\it mobile} particles identified as the $5 \%$ most mobile particles at that time are defined to belong to the same cluster if they are within a distance $1.4\sigma_{AA}$ of each other (We consider distances at the time the clusters are identified, but do not find significant changes in the distribution of clusters if the separation of particles at the initial time is considered instead). The average of the sizes of these clusters is normalised by the average cluster size when the subset labeled as mobile is randomly chosen in the initial configuration. In practice, we compute the normalisation based on labeling particles as mobile considering their displacements in the first integration step at the time origin.

We next consider string-like cooperatively moving particles. Strings have been identified as groups of mobile particles such that the position of one of the particles is occupied by another particle at a later time \cite{donati1998stringlike}. After we identify mobile particles at a time $t$, we check if the $i^{th}$ mobile particle has been replaced by the $j^{th}$ mobile particle within the radius $\delta$ over the interval $t$. If so, those two particles are considered to form a string. Here, two particles are identified as belonging to the same string if $|r_j(0)-r_i(t)|] < 0.6 \sigma_{AA}$. It has been observed that for a given $i$, multiple other particles may satisfy such a criterion, in which case, the particle $j$ which has the minimum distance $|r_j(0)-r_i(t)|]$ is identified as the particle that replaces $i$. In computing the average string length, we include particles that are not connected with any other as strings of length $1$, so that at very short and very long times, the average length of a string tends to a value of $1$.

Fig. \ref{mobstringtevolve} (a) shows the time dependence of the mean size of mobile clusters, which exhibit a maximum at a characteristic time $t_{n_M}^{peak}$. The mean length of strings is shown as a function of time in Fig. \ref{mobstringtevolve} (b), which exhibits a maximum at a characteristic time $t_L^{peak}$. These times are plotted against the time $t^{*}$ at which $\alpha_2(t)$ is maximum, in Fig. \ref{mobstringtevolve} (c). Consistently with observations in \cite{starr2013relationship}, these times are seen to be essentially the same (but see below for further discussion on this point).

\subsection{Summary of various timescales}

\begin{figure}[t]
\centering
\includegraphics[scale=.43]{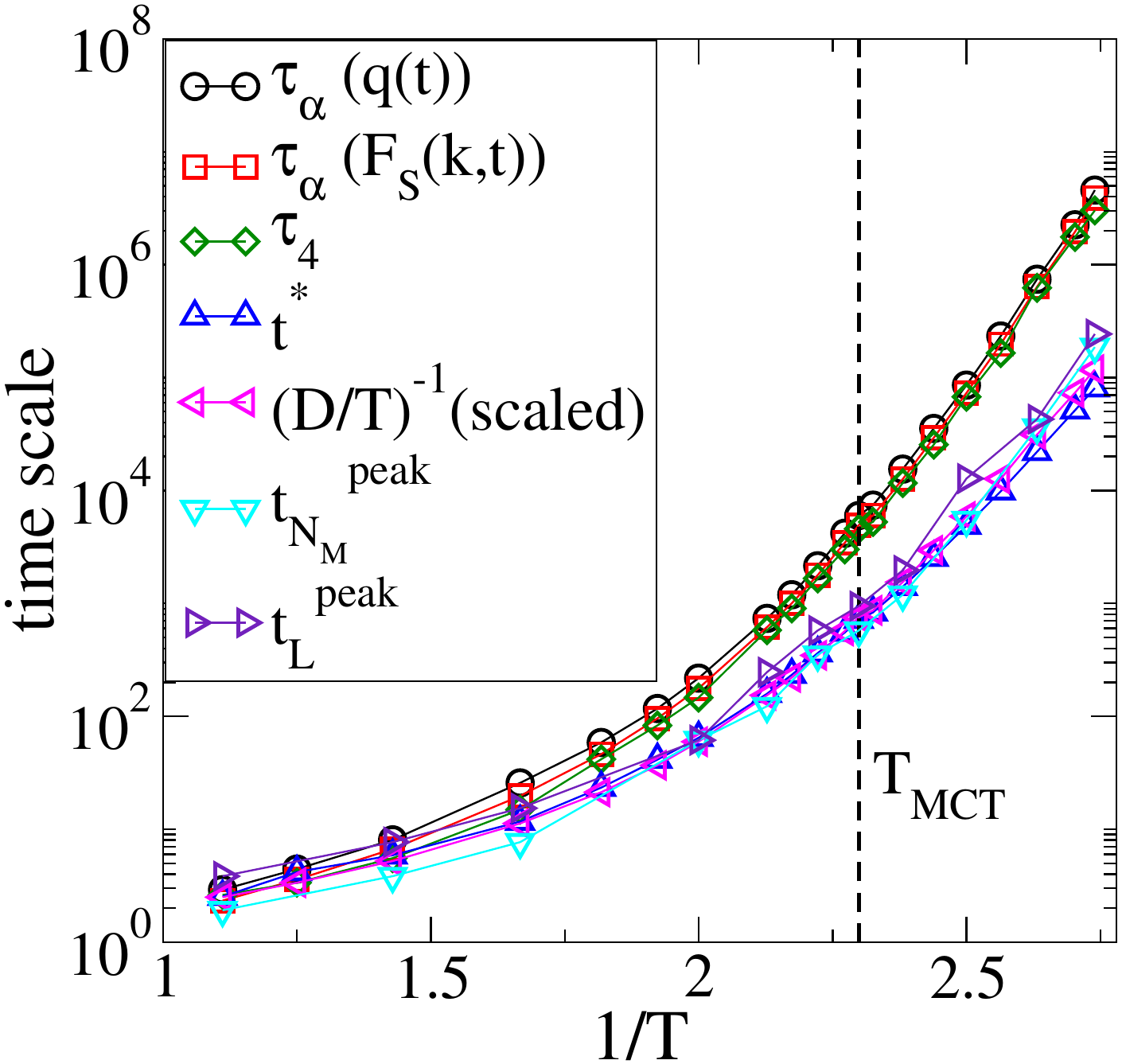}
\caption{Different time scales, $\tau_{\alpha}$ (from $q(t)$ and $F_s(k,t)$), $\tau_4$, $t^{*}$, $\left(D/T\right)^{-1}$, $t_{n_M}^{peak}$, $t_L^{peak}$, plotted against $1/T$ in an Arrhenius plot. All time scales show a crossover to Arrhenius behavior at low temperatures. They fall in to two groups: $\tau_{\alpha}$ and  $\tau_4$ are larger and have stronger $T$ dependence, whereas $t^{*}$, $\left(D/T\right)^{-1}$, $t_{n_M}^{peak}$, and $t_L^{peak}$, are smaller, and show weaker $T$ dependence. 
}
\label{allscale}
\end{figure}

We have reported above on different characteristic times, defined with respect to structural relaxation and dynamical heterogeneity. We summarise the temperature dependence of these time scales here and compare them with each other. Fig. \ref{allscale} shows an Arrhenius plot of the alpha relaxation time $\tau_{\alpha}$ (from $q(t)$ and $F_s(k,t)$), the diffusion time scale $\left(D/T\right)^{-1}$, the time at which $\chi_4$, $\alpha_2$, the size of mobile particles $n_M$, string length $L$ show maximum values, which are, respectively, $\tau_4$,  $t^{*}$, $t_{n_M}^{peak}$, and $t_L^{peak}$. The time scales shown fall into two groups: The time scales 
 $\tau_{\alpha}$ and $\tau_4$ are essentially the same, as previously observed \cite{glotzer2000time,karmakar2009growing,karmakar2014growing,karmakar2016length,adhikari2021spatial}. The remaining time scales, $\left(D/T\right)^{-1}$ (which has been scaled to match the magnitude of the others at one reference temperature), $t^{*}$, $t_{n_M}^{peak}$, and $t_L^{peak}$, also exhibit the same temperature dependence, which is milder than that of  $\tau_{\alpha}$\cite{starr2013relationship,adhikari2021spatial}. We note that recent work on a metallic glass former indicates that $t_{n_M}^{peak}$ is larger than $t^{*} \sim t_L^{peak}$ and is equal to the time scale associated with the Johari-Goldstein process \cite{zhang2021dynamic}, although all these time scales exhibit a milder temperature dependence than the alpha relaxation time $\tau_{\alpha}$. Such a distinction is not apparent from our present results. 

The decoupling of the diffusion time scale and the alpha relaxation time scale have been investigated extensively (\textmark{\cite{Rossler1990,thirumalai1993activated,stillinger1994translation,tarjus1995breakdown,andreozzi1996study,cicerone1996enhanced,douglas1998obstruction,berthier2004length,berthier2004time,kim2005breakdown,chong2009coupling,sengupta2013breakdown,charbonneau2014hopping,parmar2017length,adhikari2021spatial}} and references therein), in the context of the breakdown of the Stokes-Einstein relation, employing $\tau_{\alpha}$ as being proportional to the viscosity \cite{sengupta2013breakdown}. We consider the breakdown of the Stokes-Einstein relation in order to investigate whether it reveals any indication of the crossover in dynamics around $T_{MCT}$, as reported in \cite{zhang2021dynamic}. We plot the diffusion coefficient $D$ (obtained from mean squared displacement plots shown in \ref{appendixC}) {\it vs.} $\tau_{\alpha}$ in Fig. \ref{SER}. The results clearly display a breakdown of the Stokes-Einstein relation, consistently with previous results in \cite{sengupta2013breakdown,parmar2017length}, but do not exhibit any marked change in behavior around $T_{MCT}$. The breakdown exponent $\xi$ we obtain by fitting the $D$ values to the form $D \propto \tau_{\alpha}^{-\xi^{SE}}$ is $\xi^{SE} = 0.81$, which is consistent with values in the range $0.78$ to $0.83$ previously reported \cite{sengupta2013breakdown,parmar2017length}.

\begin{figure}[t]
\centering
\includegraphics[scale=.43]{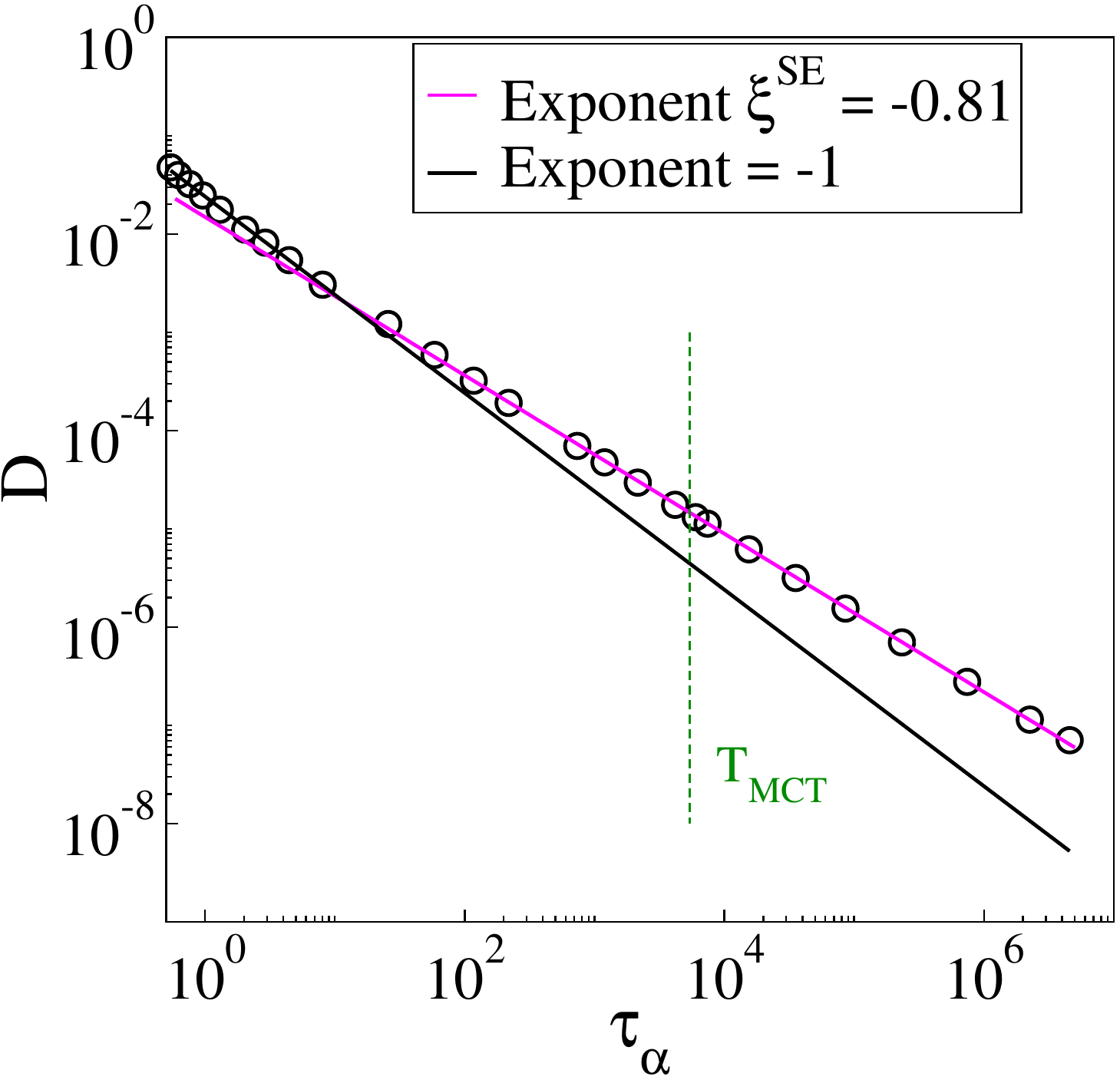}
\caption{The diffusion coefficient $D$ is shown against $\tau_{\alpha}$ exhibiting the breakdown of the Stokes-Einstein relation. A fit to high temperature data with exponent $-1$ is shown for reference. Results below $T = 0.8$ exhibit a best fit exponent of $-0.81$ and the behavior in this regime does not show any indication of a crossover around $T_{MCT}$. 
} 
\label{SER}
\end{figure}


\subsection{Morphology of correlated rearrangements}

\begin{figure*}[t]
\centering
\includegraphics[scale=.45]{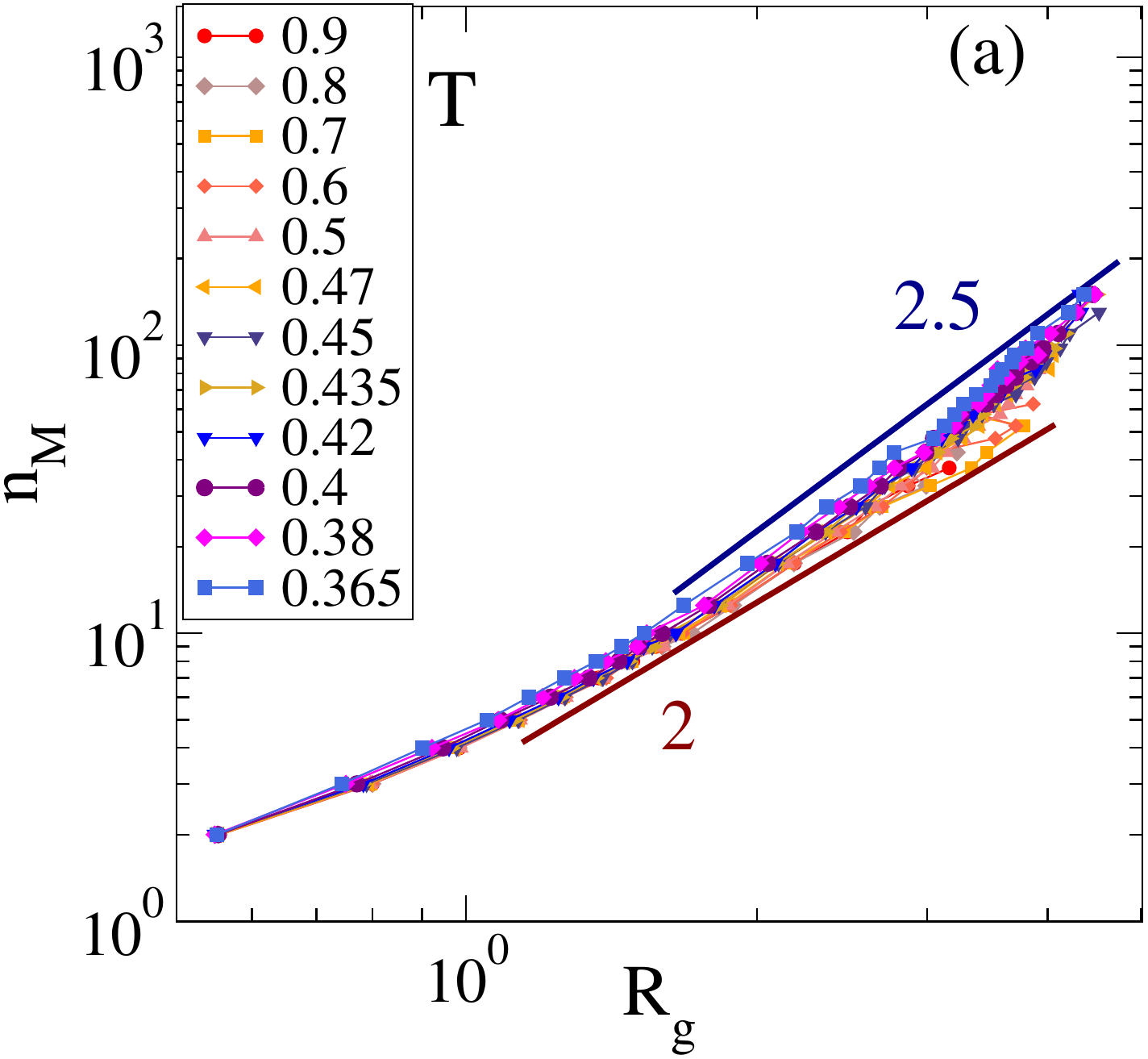}
\includegraphics[scale=.45]{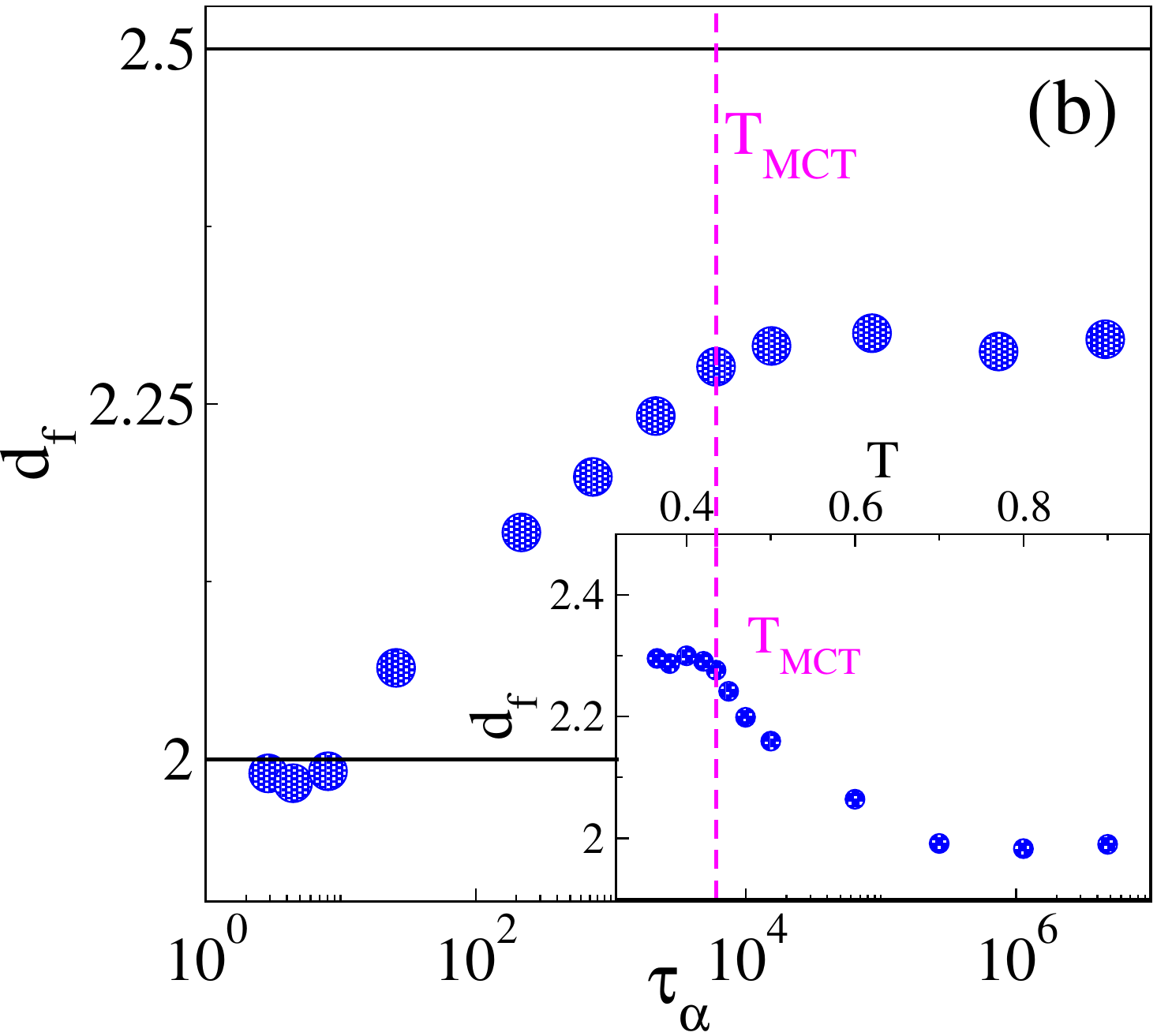}
\caption{The fractal dimension of mobile clusters: (a) The dependence of the size of the clusters on the radius of gyration, for different temperatures. (b) The fractal dimensions $d_f$ shown as a function of $\tau_{\alpha}$ or temperature (inset) reveal a marked change as $T_{MCT}$ is traversed. The horizontal lines mark $d_f = 2.0$ and $d_f = 2.5$ respectively, for reference. 
}
\label{rg}
\end{figure*}

In addition to the sizes of rearranging regions, there has been considerable interest in analysing the morphology of the correlated rearranging regions. In addition to investigations of string-like cooperative motion already mentioned \cite{donati1998stringlike}, the observation of string-like rearrangements at temperatures accessed in computer simulations had led to theoretical analysis within the framework of the random first order transition theory (RFOT) \cite{stevenson2006shapes}, leading to the prediction of a crossover of rearranging regions from compact to fractal morphology at the dynamical transition temperature, identified with the mode coupling transition. In \cite{starr2013relationship}, the fractal dimensions of mobile, immobile clusters and strings were analysed. It was found that mobile and immobile clusters exhibited a change in the fractal dimension from $d_f \sim 2$ to $d_f \sim 2.5$ as the temperature was lowered, and strings exhibited a change from $d_f \sim 5/3$ to $d_f \sim 2$. For mobile and immobile clusters, the fractal dimensions found were described as being in the range of the $d_f$ value of $2$ for lattice animals, to $d_f = 2.5$ observed for branched polymers with screened excluded volume interactions. Similarly, $d_f = 5/3$ is the fractal dimension for self-avoiding walks, and $d_f = 2$ is the fractal dimension for self-avoiding walks with screened excluded volume interactions. We consider here the fractal dimensions for mobile clusters, following the results in \cite{starr2013relationship}, to investigate whether a crossover in the cluster morphology is observed that accompanies the dynamical crossover.


For a cluster of size $n$, we may define the fractal dimension $d_f$ from its dependence of the radius of gyration, 
\begin{equation}
n\sim R_g^{d_f}
\label{dfeq} 
\end{equation}
where,
\begin{equation}
R_g^2=\frac{1}{2 n^2}\sum_{i,j}(r_i-r_j)^2
\end{equation}
The results obtained at different temperatures of the dependence of the cluster size on the radius of gyration are shown in Fig. \ref{rg} (a) for mobile clusters. As observed in \cite{starr2013relationship}, the cluster size does not depend on $R_g$ with a single power law exponent, which indicates that as the clusters grow larger, their morphology changes. We consider clusters of size $> 5$ and at each temperature, and obtain the fractal dimension with a single best fit to the form in Eq. \ref{dfeq}. Such a procedure will provide an underestimate of the fractal dimension of the largest clusters, but alternate procedures lead to comparable estimates. The fractal dimensions so obtained are shown in Fig. \ref{rg} (b), both as a function of temperature (inset) and of $\tau_{\alpha}$. Remarkably, we find that a clear crossover is observed for the fractal dimension in each case, showing that the dynamical transition we observe is indeed accompanies by a change in the morphology of correlated rearranging regions. Although Fig. \ref{rg} (a) suggests that the largest clusters approach a fractal dimension of $d_f = 2.5$, our numerical estimates saturate at a lower value.

\subsection{Thermodynamics}

\begin{figure*}[t]
\centering
\includegraphics[scale=0.45]{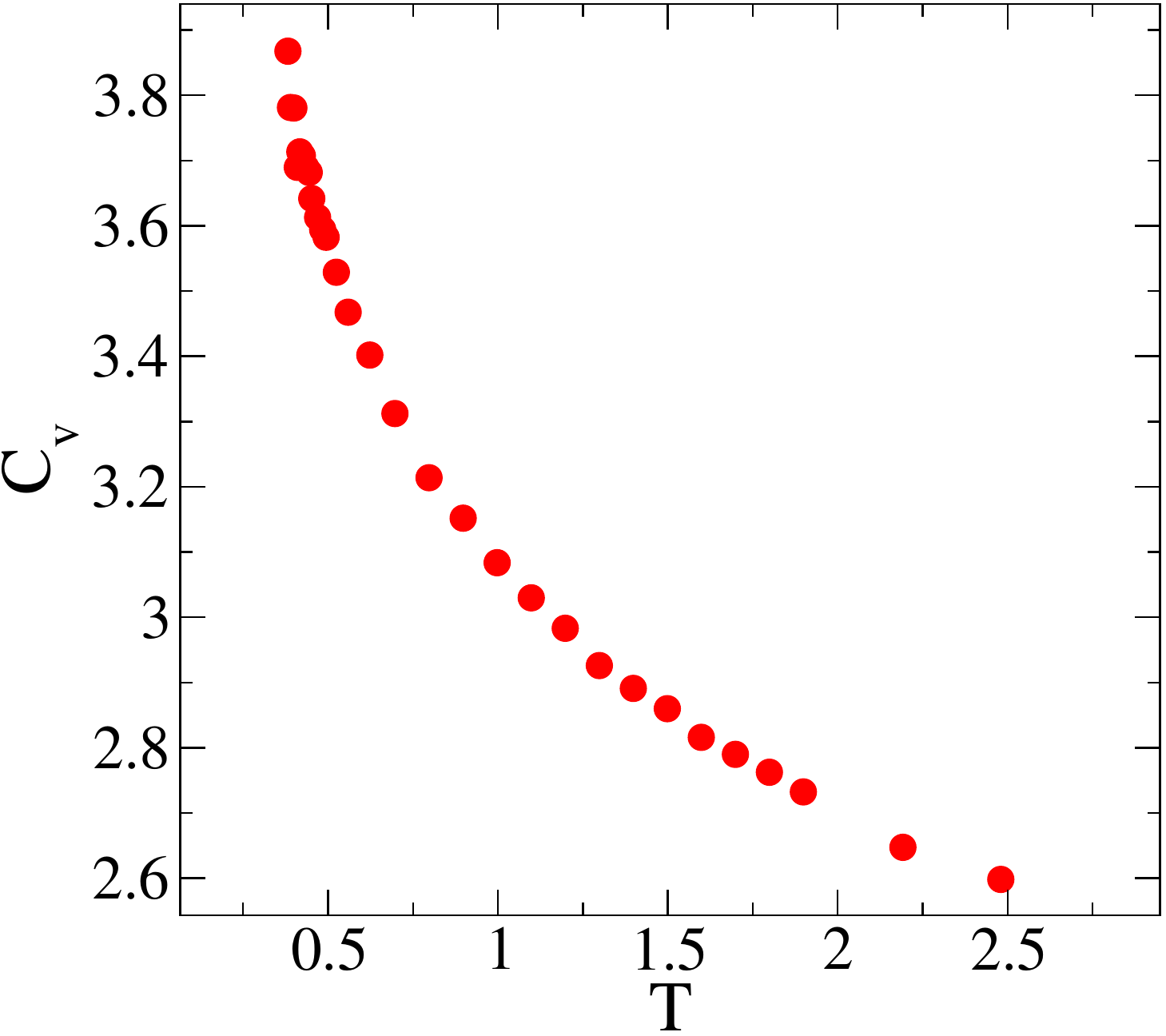}
\includegraphics[scale=0.45]{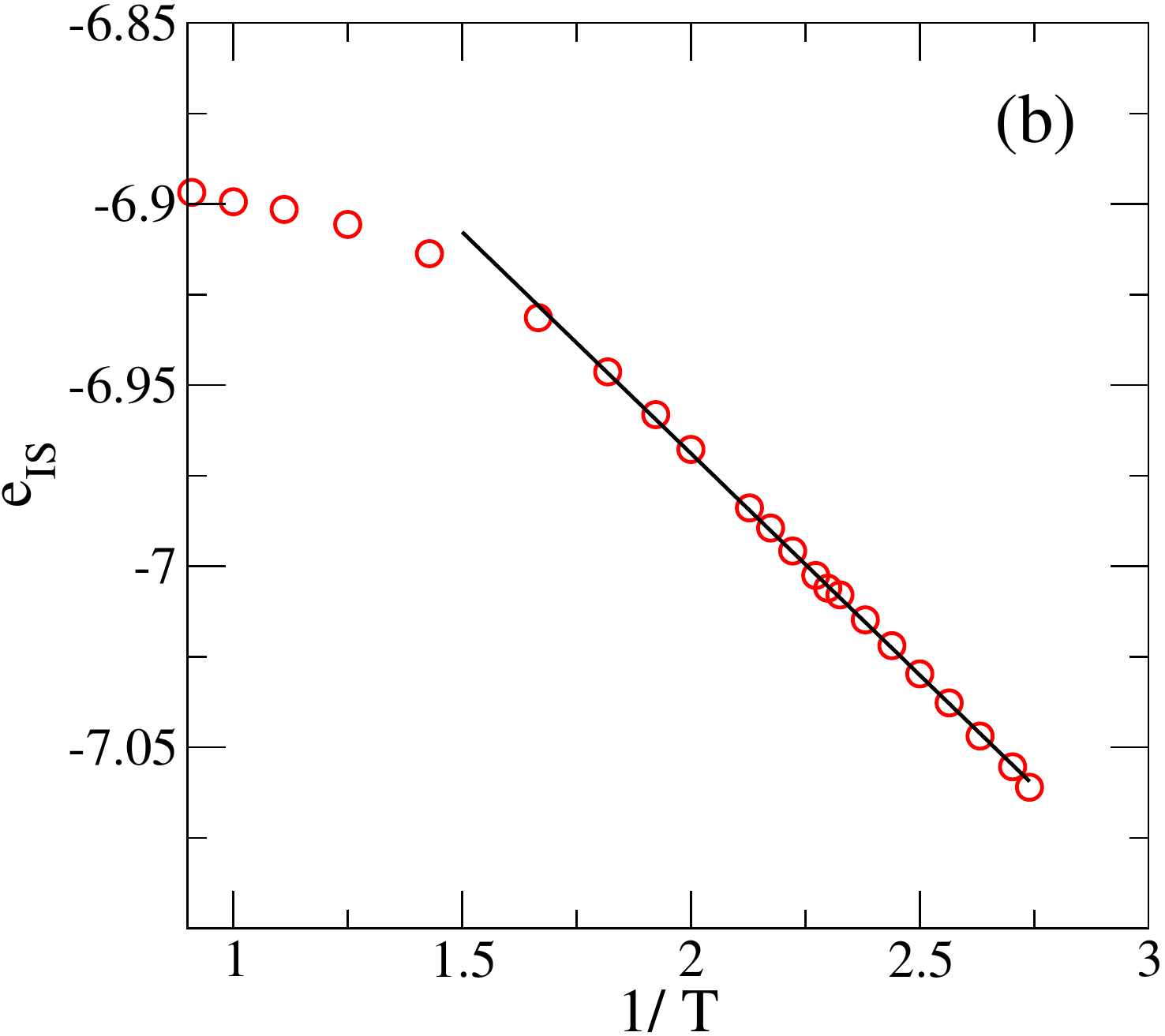}
\caption{\textmark{(a)} The specific heat $C_v$, plotted against temperature, shows a monotonic increase as the temperature is lowered. 
(b) The inherent structure energy $e_{IS}$ plotted against inverse temperature, shows a $1/T$ dependence at all temperatures below $T \sim 0.7$. 
}
\label{cv}
\end{figure*}

We consider next the thermodynamic aspects of the dynamical crossover observed. As mentioned in the introduction, the fragile to strong crossover has been associated with the presence of a specific heat maximum. We thus first consider the constant volume specific heat $C_v$ obtained by differentiating numerically the internal energy with temperature. The resulting specific heat, shown in Fig. \ref{cv} (a), displays a monotonic increase as the temperature is decreased, consistently with \cite{coslovich2018dynamic}. We note that the presence of a specific heat maximum has been reported in \cite{depablo2004cpmax,flenner2006} for the model studied here and a similar binary mixture glass former in a similar temperature range as the lowest temperatures we study, which may have arisen either as a result of a lack of equilibration or system size effects. Such a maximum has also been reported in a metallic glass model which exhibits a more marked fragile to strong crossover \cite{zhang2021dynamic}. Our results indicate that the dynamical crossover we observe is not related to the presence of a specific heat maximum, at least in the temperature range investigated. 

A related quantity that has been investigated in the context of the fragile to strong crossover ({\it e. g.}, in the context of silica \cite{saika2004,heuersilica}) is the average energy of local energy minima, or inherent structures, $e_{IS}$, as a function of temperature. The average inherent structure energy displays a $1/T$ dependence for, {\it e. g.}, KA-BMLJ \cite{Sastry2000prl,Sastry2001}, but displays deviations for liquids displaying a fragile to strong crossover. The inherent structure energies shown in Fig. \ref{cv} (b) do indeed show a $1/T$ temperature dependence below $T \sim 0.7$, and more importantly, do not show any indication of a deviation from the $1/T$ behavior down to the lowest temperatures investigated. 

\begin{figure*}[t]
\centering
\includegraphics[scale=.4]{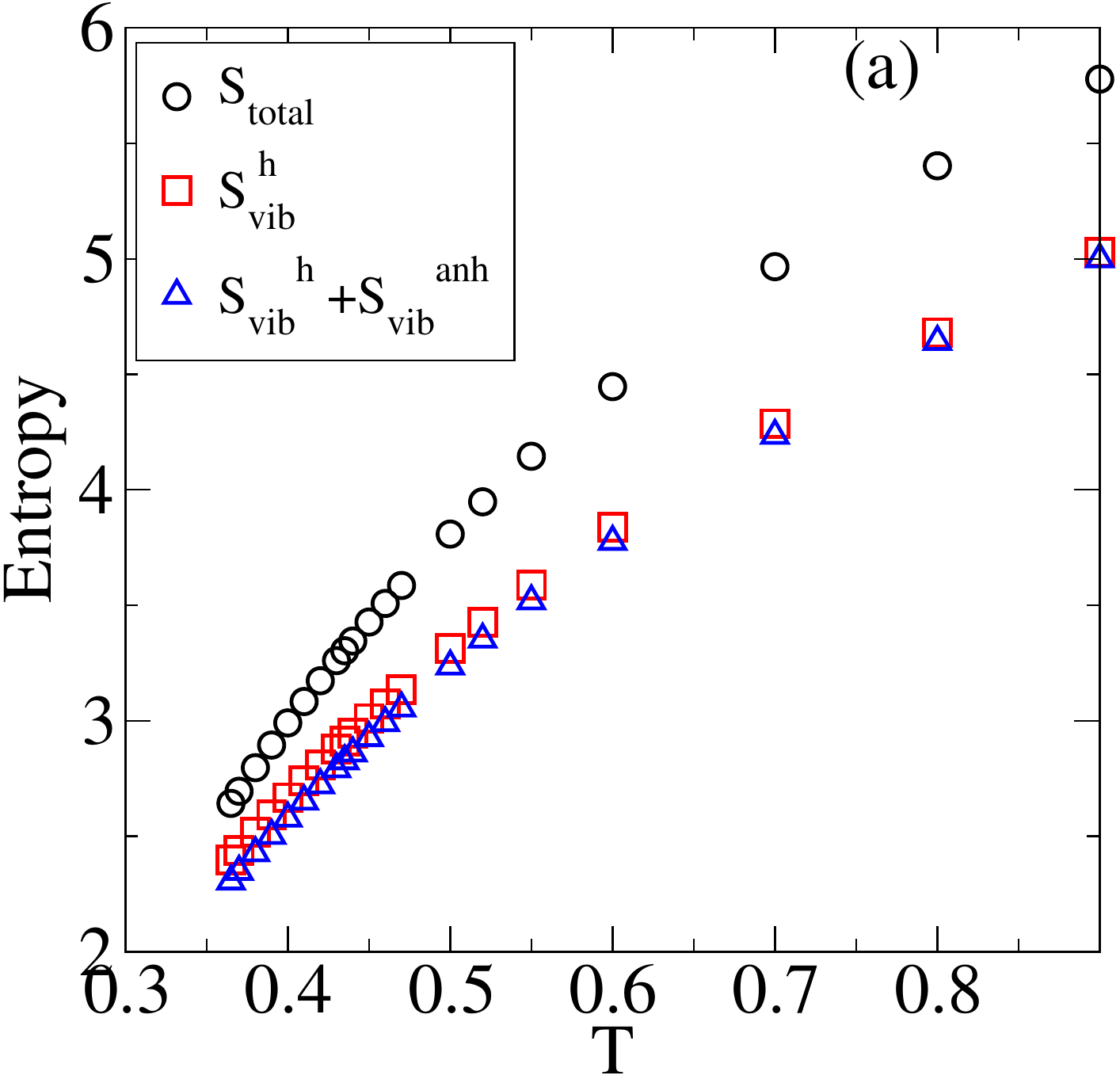}
\includegraphics[scale=.4]{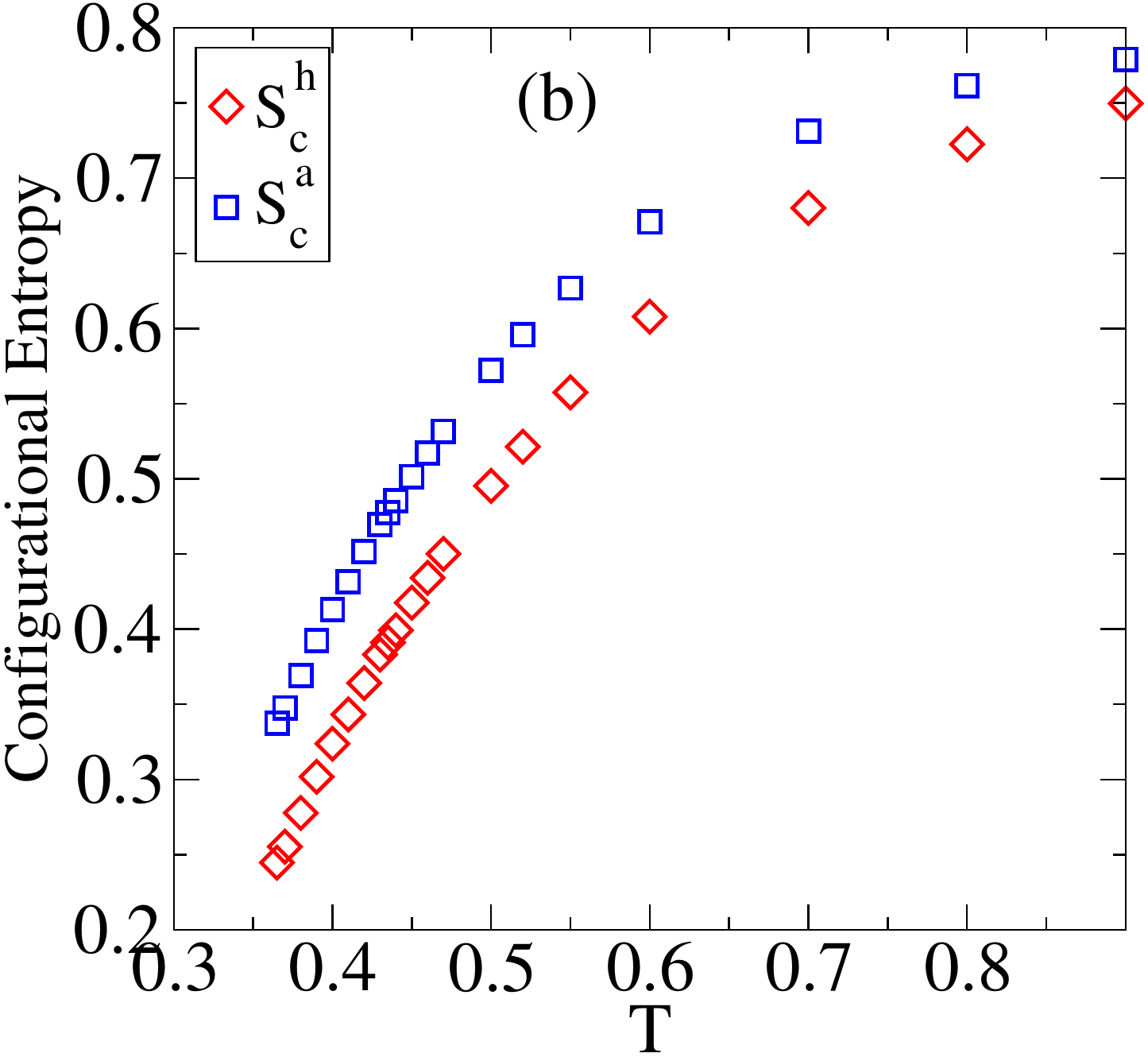}
\caption{\textmark{(a)} The vibrational entropy along with the total entropy is shown as a function of the temperature. The vibrational entropy computed according to the harmonic approximation ($S_{vib}^h$) as well as by including the anharmonic correction ($S_{vib}^{anh}$), are shown. \textmark{(b)} The configurational entropy is shown as a function of the temperature, obtained by subtracting from the total entropy the vibrational entropy without ($S_c^h$) and with ($S_c^a$) the anharmonic correction.
}
\label{AGS}
\end{figure*}

\begin{figure*}[t]
\centering
\includegraphics[scale=.4]{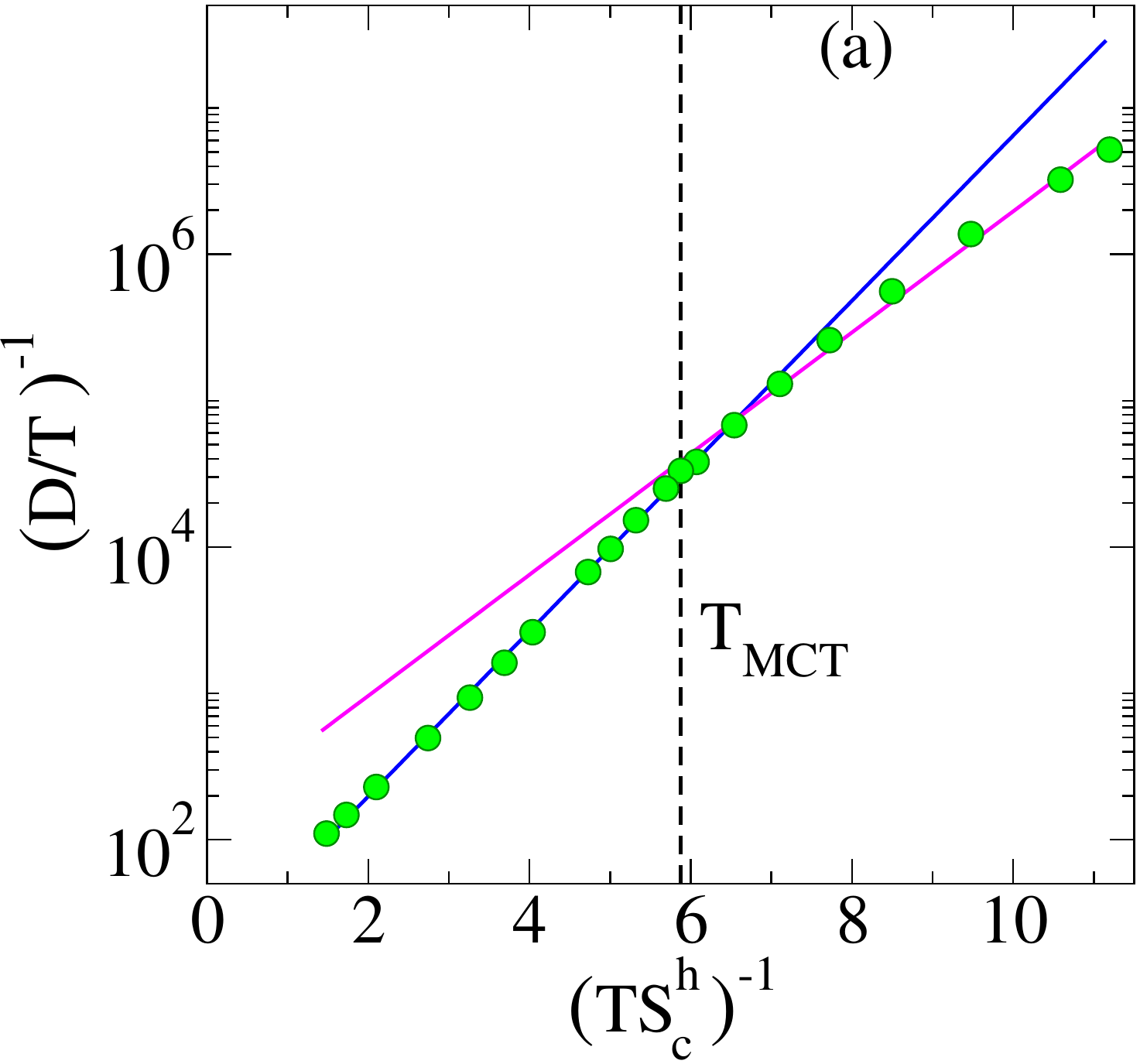}
\includegraphics[scale=.4]{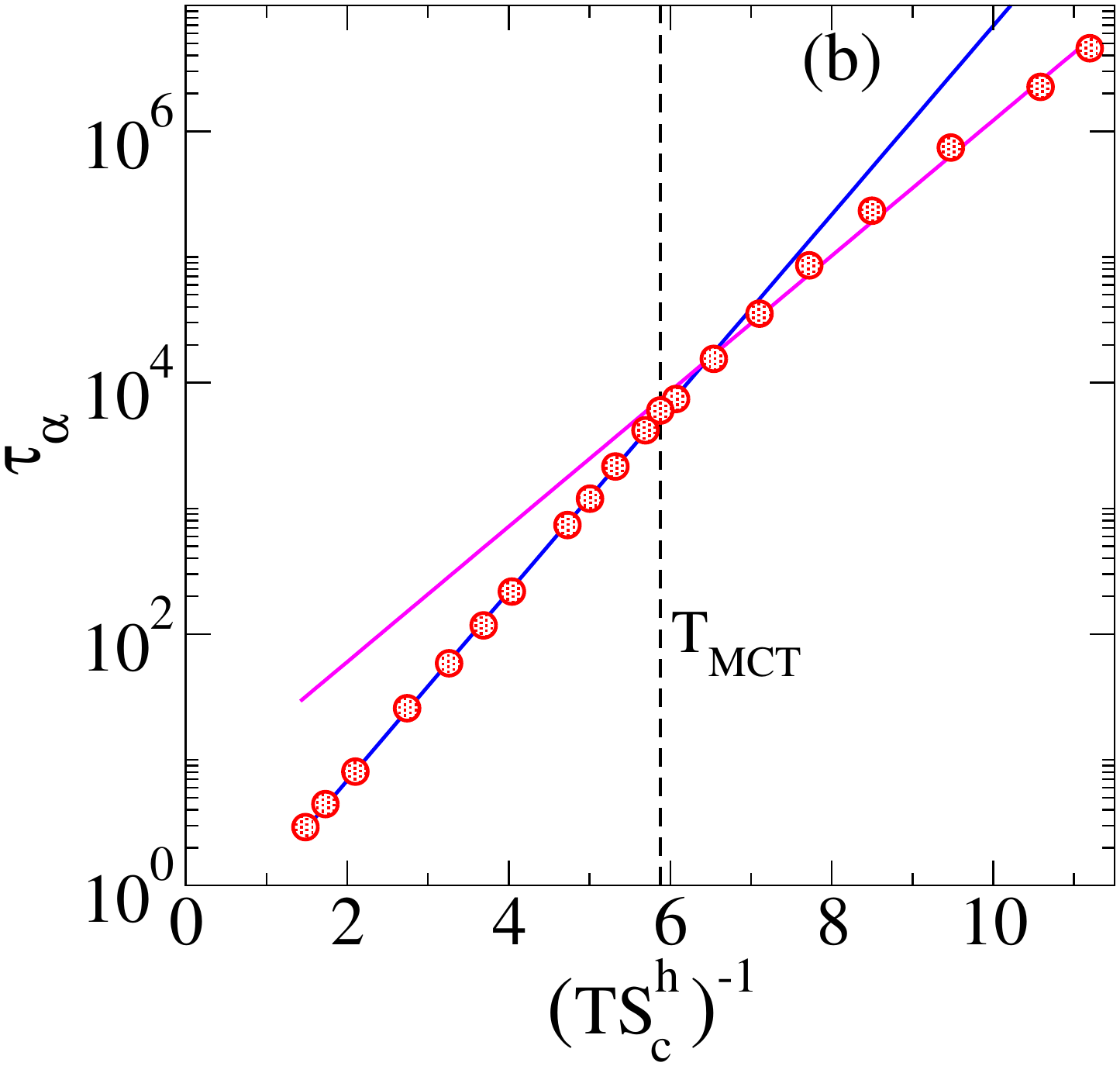}
\caption{Adam-Gibbs plots of diffusion time $\left(D/T\right)^{-1}$ and $\tau_{\alpha_A}$ employing configurational entropies $S_c^{h}$ obtained with the harmonic approximation to the vibrational entropies. A change in slope is observed around $T_{MCT}$, indicated by the vertical line. The fitted Adam-Gibbs coefficient $A$ for diffusion times is \textmark{$A_D = 1.30,0.95$ ($D_0=2.69,4.96$)} at  temperatures above and below $T_{MCT}$. Similarly, for $\tau_{\alpha}$, \textmark{$A_{\tau} = 1.73,1.24$ (${\tau}_0=-1.53,1.62$)} at temperatures above and below $T_{MCT}$. \textmark{The blue lines are fit lines for the data in the range of $T > T_{MCT}$ which are extrapolated to $T < T_{MCT}$ and the magenta lines are  fit lines for the data in the range of  $T< T_{MCT}$ which are extrapolated to $T > T_{MCT}$.}}
 \label{AGSh}
\end{figure*}

\begin{figure*}[t]
\centering
\includegraphics[scale=.4]{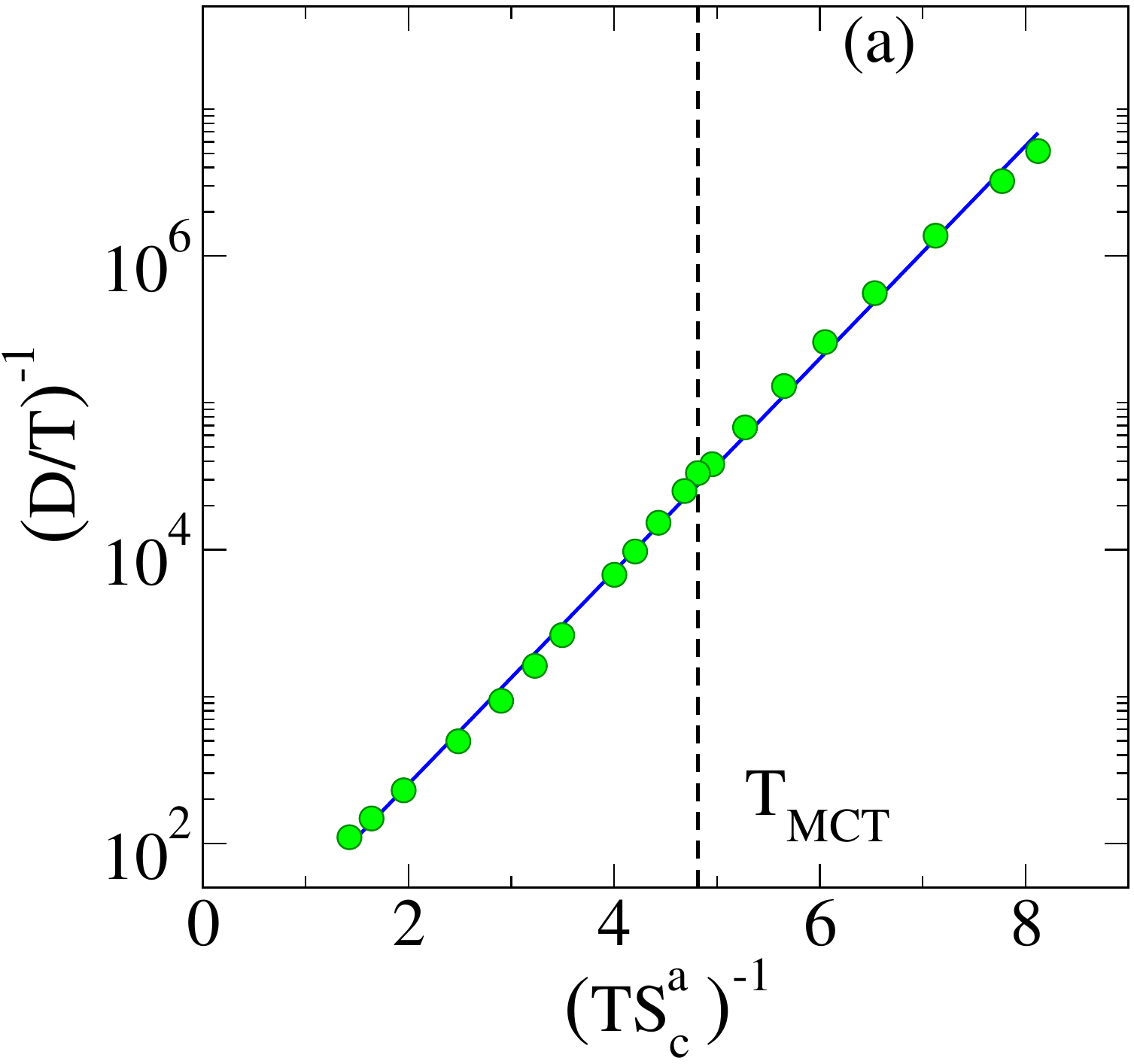}
\includegraphics[scale=.4]{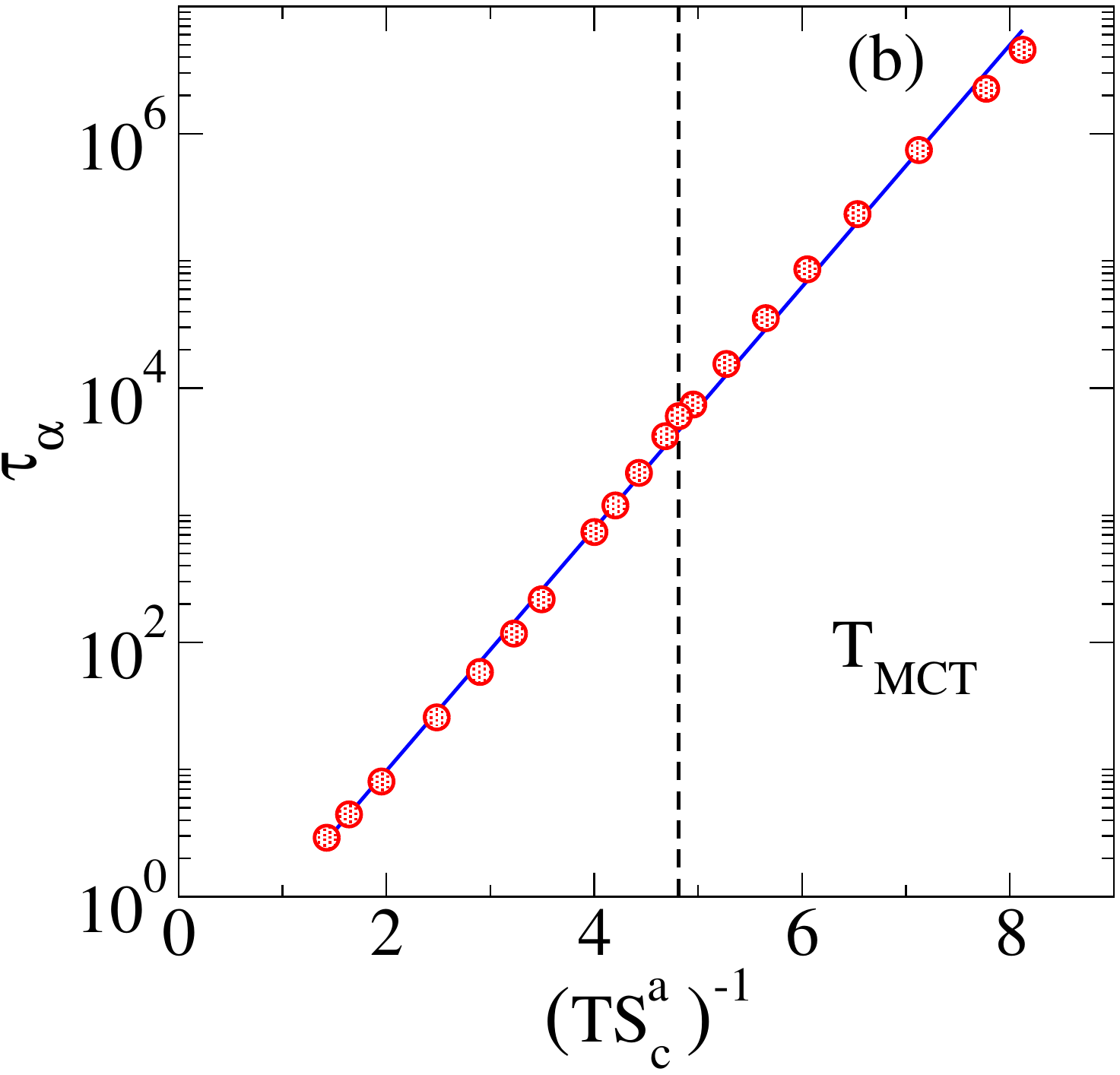}
\caption{Adam-Gibbs plots of diffusion time $\left(D/T\right)^{-1}$ and $\tau_{\alpha}$ employing configurational entropies $S_c^{a}$ obtained with the anharmonic corrections included in evaluating the vibrational entropies. The Adam-Gibbs relation is valid across  $T_{MCT}$, indicated by the vertical line.
The fitted Adam-Gibbs coefficient $A$ for diffusion times is \textmark{$A_D = 1.67$ ($D_0=2.2$)},, and similarly, for $\tau_{\alpha}$, \textmark{$A_{\tau} = 2.19$ (${\tau}_0=-2.1$ ). The blue lines are fit lines for the whole temperature range.}
 }
\label{AGSanh}
\end{figure*}


We next consider the Adam-Gibbs relation, which  relates dynamical properties such as relaxation times in glass forming liquids, to the configurational entropy, a thermodynamic quantity. The relationship can be written as 

\begin{equation}
\tau = \tau_0 ~ exp\left(\frac{A}{TS_c}\right)
\end{equation}

and has been investigated extensively, both experimentally and in computer simulations \cite{angell1998review,scala2000water,saika2001fragile,Sastry2001,Sciortino_2005,karmakar2009growing,sengupta2012adam,starr2013relationship} (where $\tau$ is a relaxation time scale, either $\tau_{\alpha}$ or the diffusion time $\left(D/T\right)^{-1}$ in most studies, including here). While several computational investigations find the Adam-Gibbs relationship to be valid, recent work\cite{ozawa2019does,ortlieb2021relaxation} raises questions about whether results from an extended range of temperatures would continue to validate the relationship. We note in particular that in \cite{ozawa2019does}, an extrapolation of relaxation times were performed using a parabolic law, which, however, has been found to be valid for the present system in only a limited temperature range, with deviations at higher temperatures and indication of deviations at lower temperatures as well \cite{coslovich2018dynamic}. Thus, accessing directly the relaxation times over a temperature window across which the character of dynamics may change provides a more reliable test. Given the crossover in dynamics in the system investigated here, by performing simulations over a much wider range of relaxation times than the previous studies mentioned, we consider whether the Adam-Gibbs relation continues to hold in this extended range of temperatures. 

The configurational entropy is calculated by subtracting the {\it vibrational} entropy associated with individual {\it glasses} (or basins of free/potential energy minima or inherent structures) from the total entropy of the liquid. Based on the observation that a harmonic approximation to the basin free energy provided a satisfactory description \cite{sciortino1999,sastry2000onset} below the onset temperature for the KA-BMLJ, the vibrational entropy has been evaluated in the harmonic approximation \cite{sciortino1999,Sastry2000prl,Sastry2001,karmakar2009growing,sengupta2012adam,parmar2017length}. However, it has been found necessary to incorporate anharmonic corrections for other systems investigated, which have been done in several ways, as reviewed in \cite{Sciortino_2005}.

Accordingly, we first compute the configurational entropy employing the harmonic approximation for the basin entropy, using the procedure outlined in \ref{appendixD}. The computed entropies, and the configurational entropies, are shown in Fig. \ref{AGS} (a) and (b) respectively. 

In Fig. \ref{AGSh} we show the Adam-Gibbs plots for $\log \tau_{\alpha}$ and the diffusion time scale $\log \left(D/T\right)^{-1}$, plotted as a function of $\left(TS_c^{h}\right)^{-1}$, where $S_c^{h}$ is the configurational entropy in the harmonic approximation. The Adam-Gibbs plots, for both $\tau_{\alpha}$ and $\left(D/T\right)^{-1}$, 
show that for temperatures below $T_{MCT}$, 
a deviation is observed from the linear behavior observed at temperatures above $T_{MCT}$. 

Although similar results have been reported in various contexts \cite{sengupta2012adam,ozawa2019does,ortlieb2021relaxation}, the observation of deviations raise questions about the relevance of anharmonic effects, since several results clearly show a change in the energy landscape topology when the mode coupling temperature is crossed \cite{broderixPRL00,leonardoPRL,tbsJCP2000,coslo,Bhaumik2019}. We thus compute the anharmonic correction to the vibrational entropy and the configurational entropy \cite{Sciortino_2005,starr2013relationship,handle2018potential,handle2018adam,BerthierJCPSc2019} as outlined in \ref{appendixD}. The vibrational entropy with anharmonic corrections are shown in Fig. \ref{AGS} (a), and the configurational entropies with anharmonic corrections are shown in  Fig. \ref{AGS} (b). Although the change in vibrational entropies appears small in  Fig. \ref{AGS} (a), inclusion of anharmonic contributions leads to a substantial change in the configurational entropies, as seen in  Fig. \ref{AGS} (b). The corresponding Adam-Gibbs plots are shown in Fig. \ref{AGSanh}. \textmark{Both $\log \tau_{\alpha}$ and $\log \left(D/T\right)^{-1}$ are linear in $\left(TS_c^{a}\right)^{-1}$ (where $S_c^{a}$ is the configurational entropy obtained after including anharmonic corrections), indicating that they obey the Adam-Gibbs relation across the entire temperature range as was found in previous studies of water \cite{handle2018potential,handle2018adam}}. We note that the results in this case are noisier, owing to the numerical errors involved in obtaining the anharmonic correction, which needs to be improved upon. However, no systematic deviation from the high temperature Adam-Gibbs behavior is seen at low temperatures\textmark{, barring small deviations at the lowest two temperatures for which the results are the least reliable}.  We thus conclude that the observed deviations when the harmonic approximation is used are an artefact of an improper accounting of the vibrational entropy. It is important to note further that, even disregarding the data at temperatures above $T_{MCT}$, the Adam-Gibbs relation is seen to be valid for roughly three decades of relaxation times below $T_{MCT}$.

\section{Discussion and conclusions}

We have described several dynamical quantities, including those that describe dynamical heterogeneity and the morphology of rearranging regions, that demonstrate a crossover in the dynamics, \textmark{when the mode coupling temperature is crossed, with relaxation times better approximated by an Arrhenius temperature dependence at lower temperatures.} Although it is tempting to describe it as a fragile to strong crossover, whether the dynamical crossover we see is a fragile to strong crossover as originally proposed by Angell \cite{angell_fragile_to_strong,ito1999thermodynamic,STARR200351} is open to question. Unlike liquids with energetically favorable tetrahedral structure (such as water, for which the fragile to strong crossover was originally proposed, and silica), the model we investigate does not display a thermodynamic signature of a change in regime in the form of a heat capacity maximum. On the other hand, several glass forming liquids typically described as fragile glass formers do display some form of a crossover at low temperatures \cite{angell1998review}, as also seen in computer simulations (for, {\it e. g.}, a model of  ortho-terphenyl\cite{rinaldi2001}). A crossover has been predicted as  a generic feature in \cite{stevenson2006shapes} within the RFOT, and in extended mode coupling theory \cite{Chong_2009mctcross}. Our results do indicate a signature in the changes in morphology of rearranging regions, although with some modifications as compared to those envisaged in \cite{stevenson2006shapes}. In seeking further a structural explanation, it will be interesting to investigate also the morphology of immobile particles, which we have not attempted in this work, in conjunction with investigations of locally preferred structures \cite{coslovich2011lfs}. Investigating the Adam-Gibbs relation, we find deviations from the high temperature conformity to the Adam-Gibbs relation at temperatures lower than $T_{MCT}$, when a harmonic approximation to the vibrational entropy is employed. However, inclusion of anharmonic contributions in estimating the vibrational entropy leads to the conclusion that the Adam-Gibbs relation is valid across the temperature range we study. A more rigorous estimation of the vibrational entropy than what we have presented here should be attempted in light of the results we present here. Another issue to consider in the present system is the possible role of finite size effects. Based on the available results, it has been argued in \cite{coslovich2018dynamic} that the observed dynamical crossover is unlikely to be a result of finite size effects. We haven't addressed this aspect any further in the present work, but with the present day computational resources, this is a question that can be more satisfactorily addressed at the present time. Our work, and related work that has been described, illustrates that exploring the nature of dynamics below the mode coupling crossover is now feasible computationally. Exploration of such low temperature dynamics should help bridge the gap between the temperature range computer simulations have been able to access in the past, and the temperature range relevant for several experimental and theoretical results.



\section*{Declaration of Competing Interest} 
The authors declare that they have no known competing financial
interests or personal relationships that could have appeared to influence
the work reported in this paper.

\section*{Acknowledgements}
This work is dedicated to the memory of C. Austen Angell, an outstanding scientist and mentor, whose seminal contributions to the understanding of glasses, water and much else, inspired the research efforts of many, including the authors of this work. We acknowledge  Monoj Adhikari, Jack Douglas, Yagyik Goswami, Jurgen Horbach, Walter Kob, Francesco Sciortino and Francis Starr for useful discussions. We thank Monoj Adhikari and Yagyik Goswami in particular for helpful interactions in preparing the manuscript. We gratefully acknowledge the Thematic Unit of Excellence on Computational Materials Science, and the National Supercomputing Mission facility (Param Yukti) at the Jawaharlal Nehru Center for Advanced Scientific Research for computational resources.  SS acknowledges support through the JC Bose Fellowship  (JBR/2020/000015) SERB, DST (India).



\appendix


\section{The overlap function} \label{appendixA}

\begin{figure}[h]
\centering
\includegraphics[scale=.5]{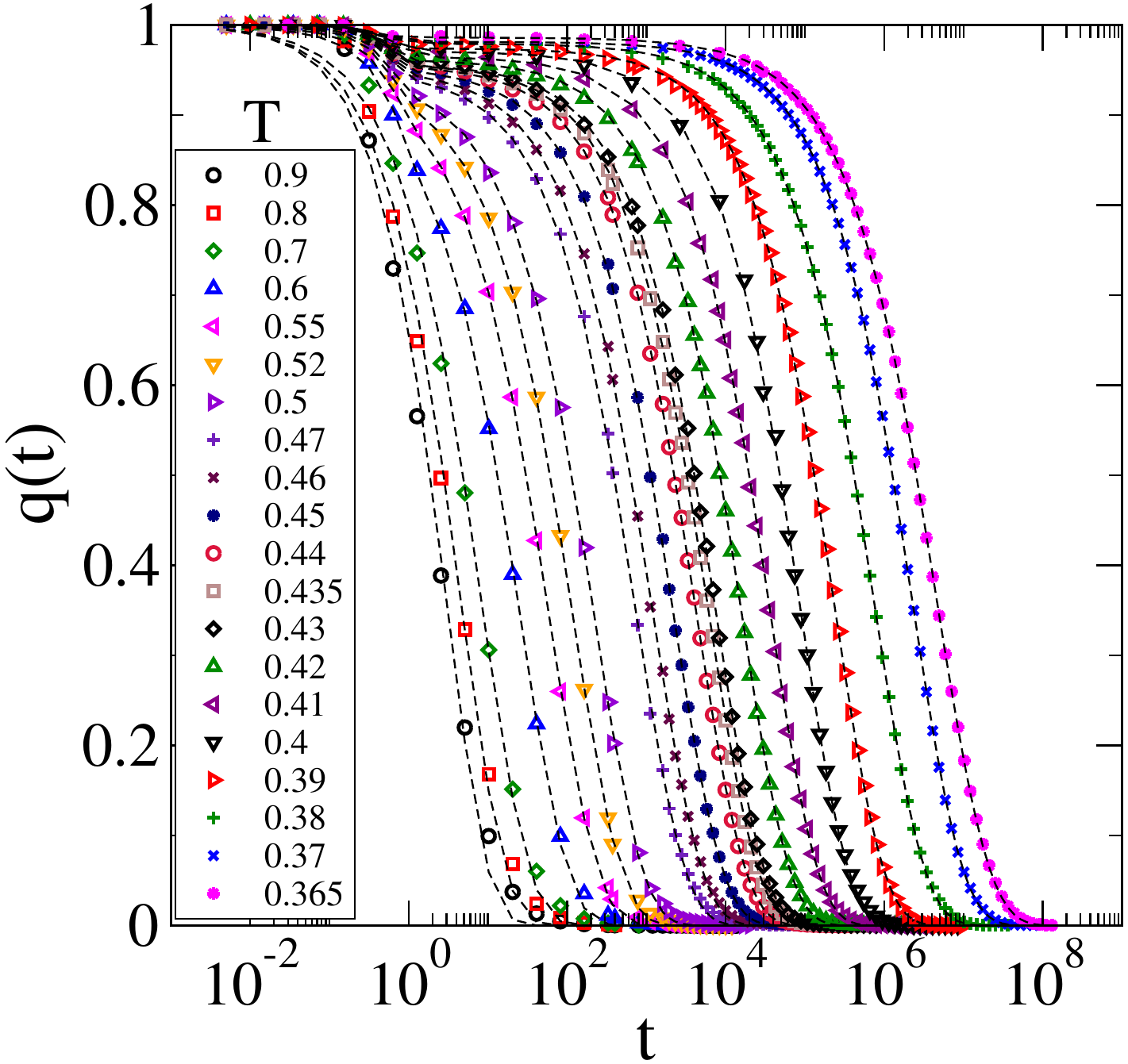}
\caption{The self part of the overlap function is shown for $A$ of particles. The dotted lines are  fits to the data.}
\label{overlap}
\end{figure}

The overlap function $q(t)$, defined as \\

\begin{eqnarray}
q(t) &=& {1 \over N} \int d \textbf{r}\rho(\textbf{r},t) \rho(\textbf{r},t+t_0) \\ \nonumber
     &=& {1 \over N} \sum_i\sum_{j}\delta(\textbf{r}_j(t_0)- \textbf{r}_i(t+t_0))
\end{eqnarray}
where $\rho(\textbf{r},t)$ is the local density of particles at position $r$ at time $t$,  can be divided into a self part and a distinct part. In the present work, we will employ the self part, $q(t)_s$, defined as 

\begin{eqnarray}
q(t)_s &=& {1 \over N} \sum_{i}\delta(\textbf{r}_i(t_0)- \textbf{r}_i(t+t_0)) \\ \nonumber
\end{eqnarray}
as a good approximation to the full overlap function. 
In simulations, the $\delta$ function is approximated by a window function $w(x)$ described below, where we further define the function considering the $A$ particles only. Thus, we consider 
\begin{equation} \label{qt-eq}
q(t) = \frac{1}{N_A}\sum_{i=1}^{N_A}w(|\textbf{r}_i(t_0)- \textbf{r}_i(t+t_0)|) \nonumber \\
\end{equation}

where  $w(x)  =  1.0$  if $x \le a$ and $ = 0$ otherwise. Here, $t_0$ is the time origin and the overlap function is calculated with an average over multiple time origins. The value of the overlap function depends on the choice of the cutoff parameter $a$.  This parameter is chosen in such a way that the particle positions separated due to vibrational motion are treated as the same. We choose $a =  0.3$, which corresponds to displacements at the plateau region of the mean squared displacement (MSD) curves, as shown in, {\it e. g.}, \cite{karmakar2016short} and used in previous literature. The $q(t)$ curves, along with fits to the form 

\begin{equation} 
q(t) = (1-f_c)exp(-(t/\tau_s))^n+f_cexp(-t/\tau_{\alpha})^{\beta_{kww}}
\end{equation} 

are shown in Fig. \ref{overlap}, and the fitted values of $\tau_{\alpha}$ and $\beta_{kww}$ are shown in the main text. 

\section{Self intermediate scattering function} \label{appendixB}

\begin{figure} 
\centering
\includegraphics[scale=.35]{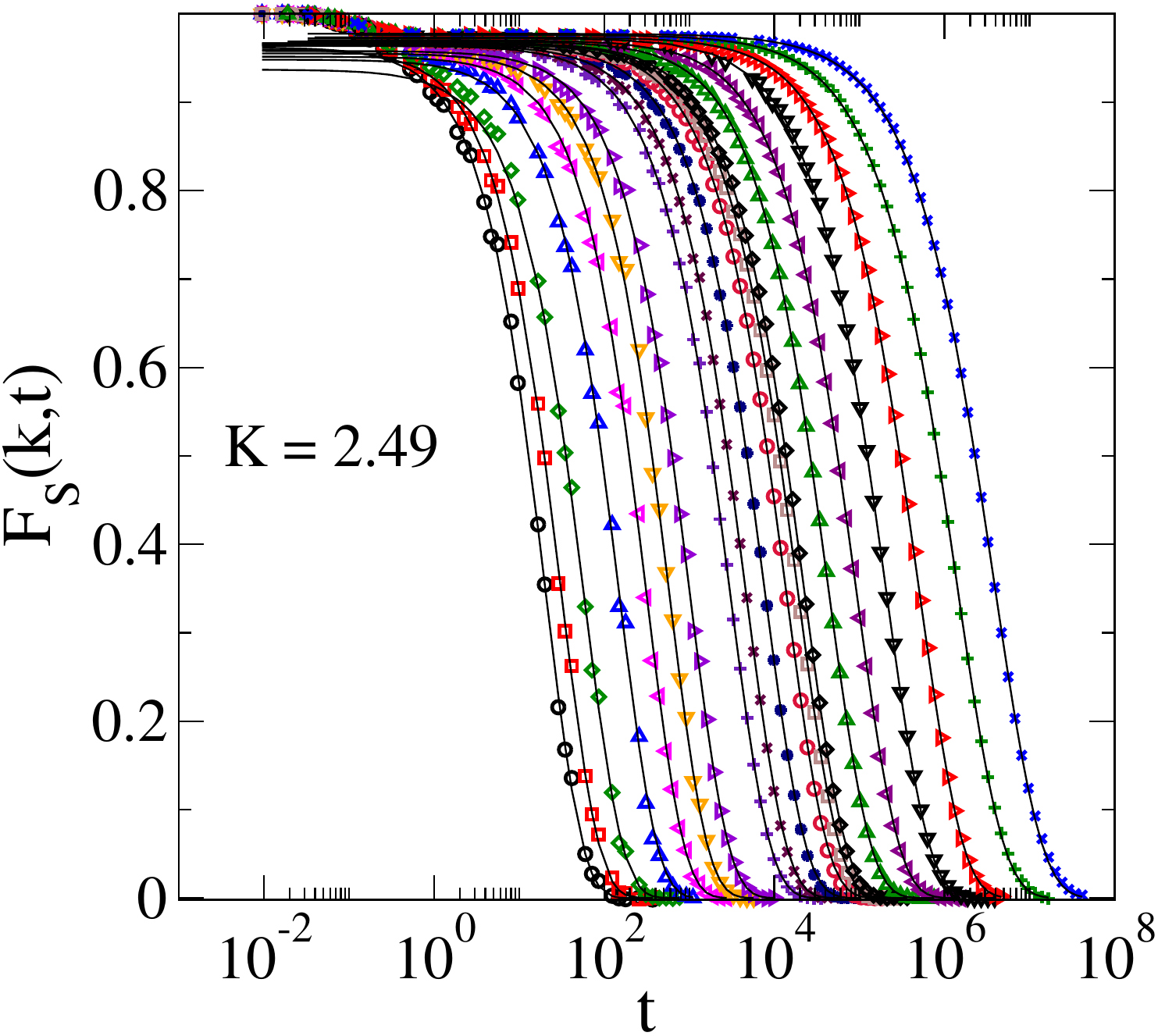}\\
\includegraphics[scale=.35]{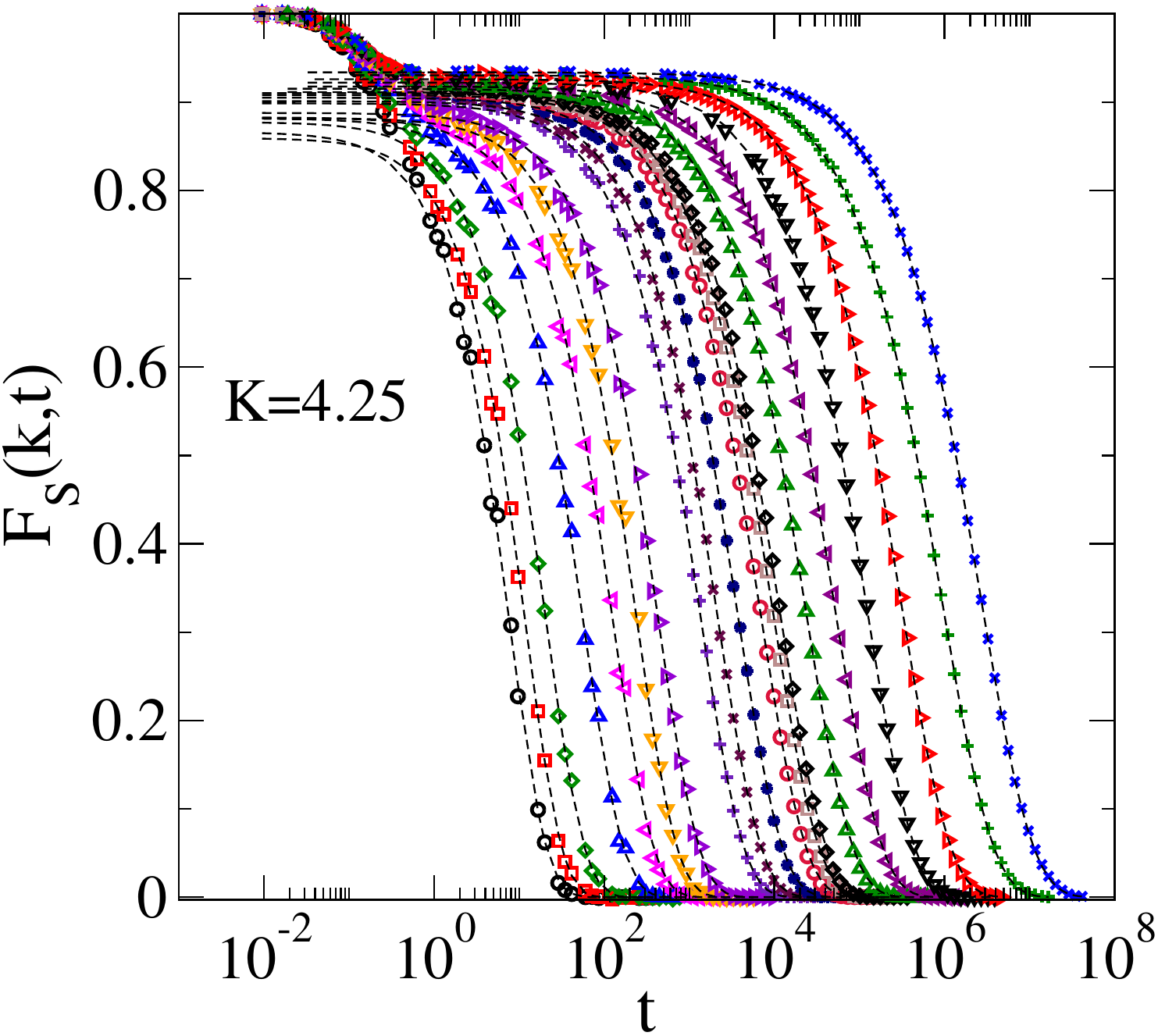}\\
\includegraphics[scale=.35]{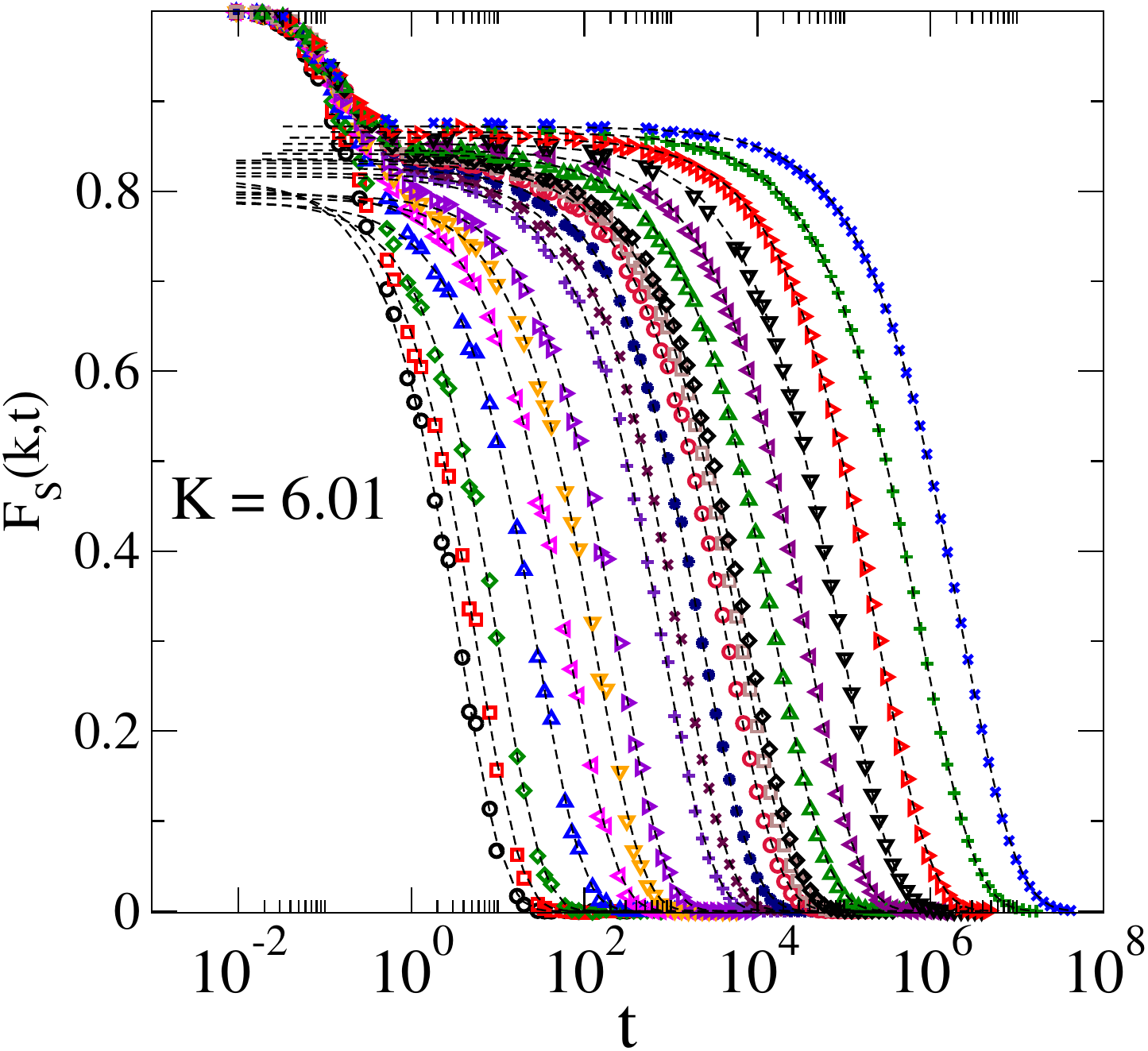}\\
\includegraphics[scale=.35]{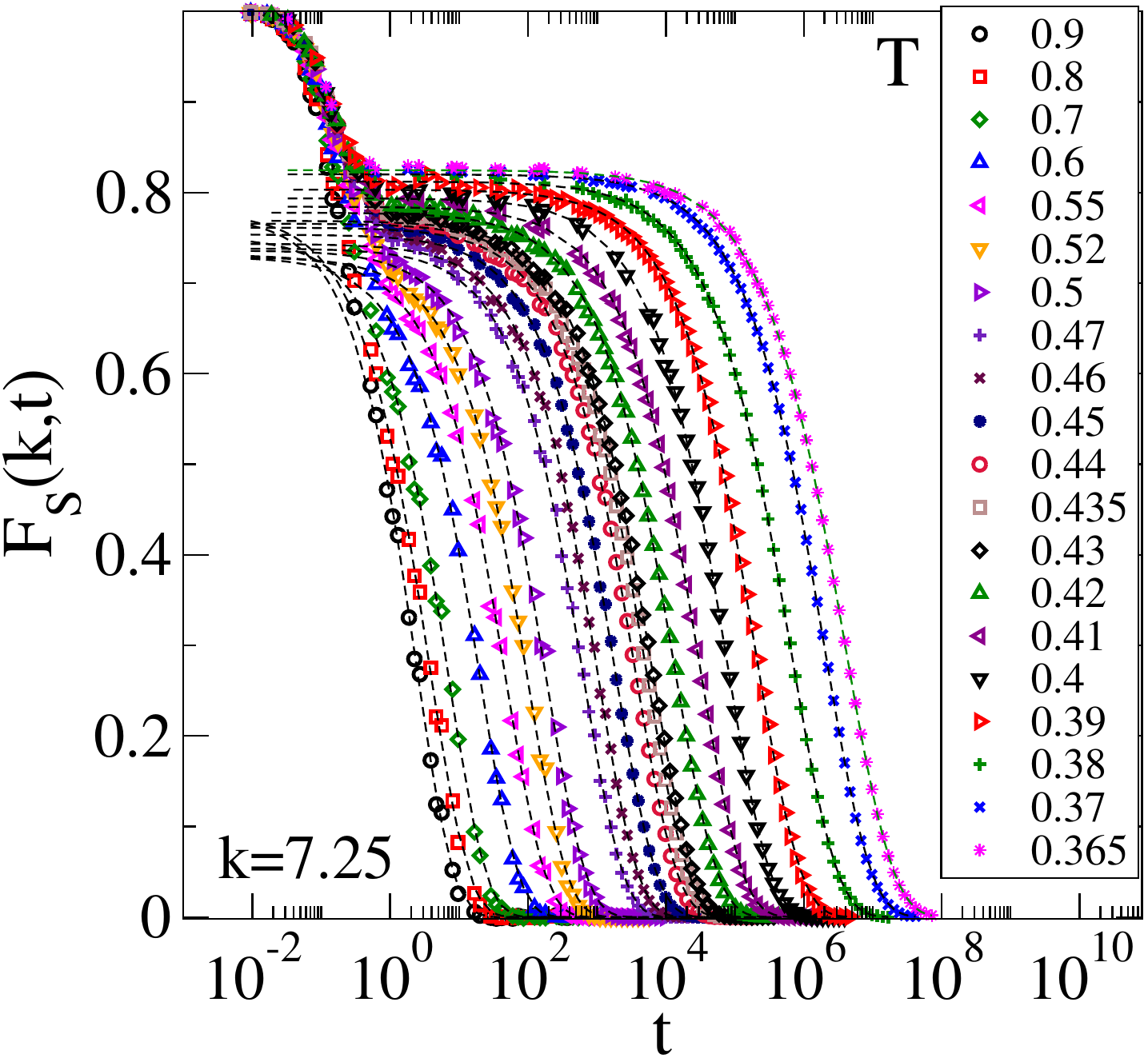}\\
\includegraphics[scale=.35]{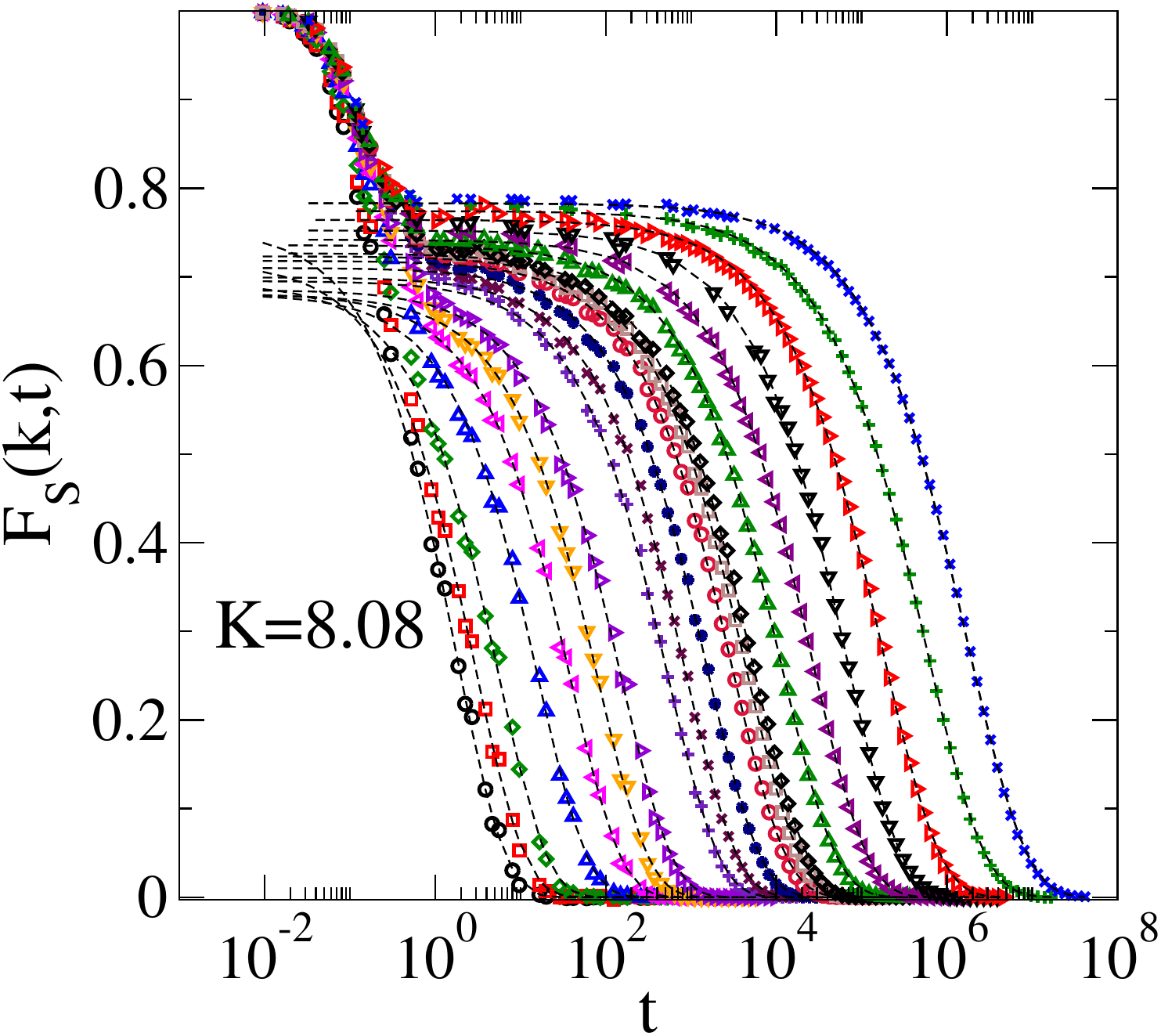}\\
\caption{The self part of the intermediate scattering function $F_s(k,t)$ is shown for a range of $k$ values. The dotted lines are fits to the data.}
\label{fsktdat}
\end{figure}

\begin{figure}[h!]
\centering
\includegraphics[scale=0.45]{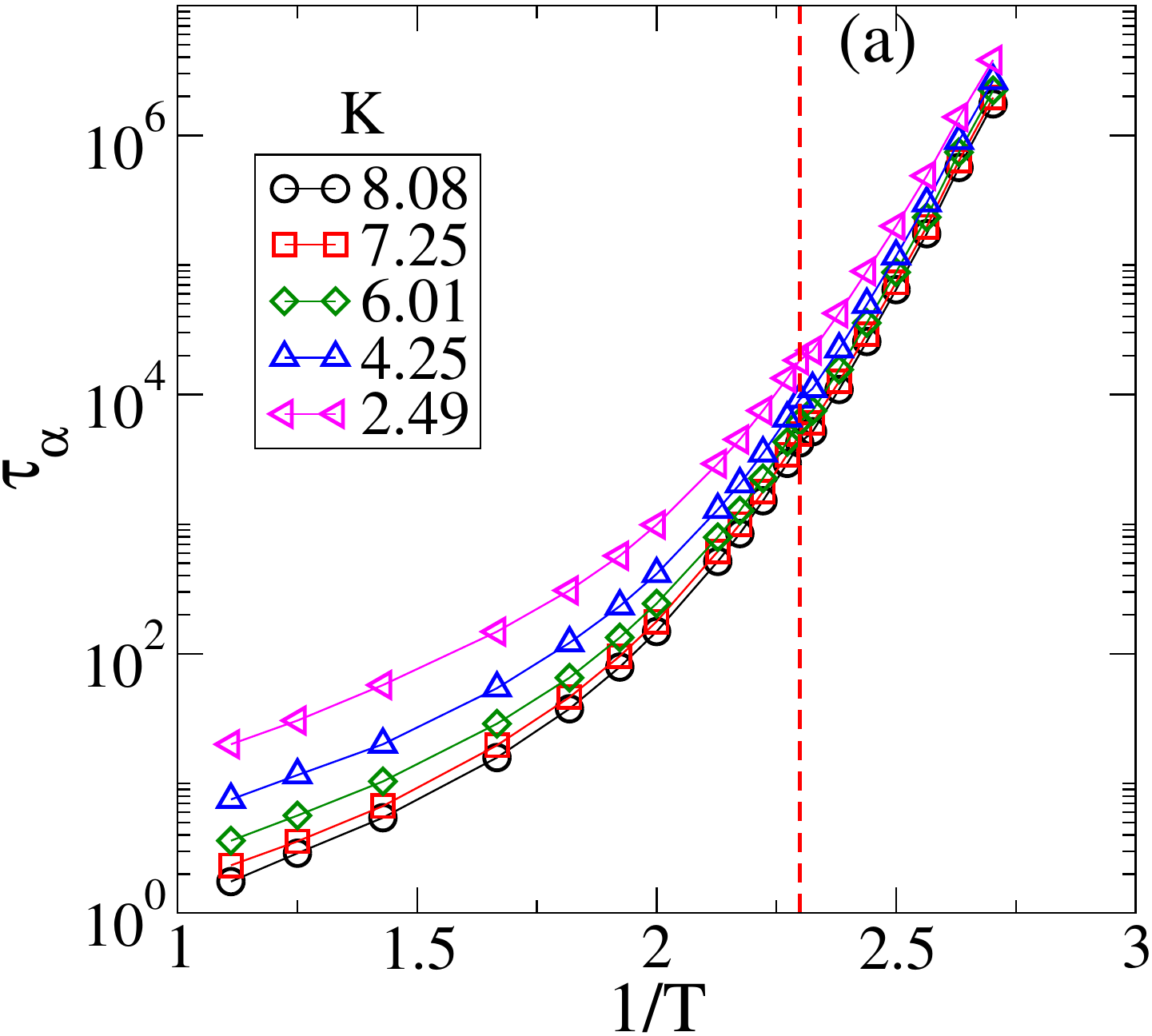}
\includegraphics[scale=0.45]{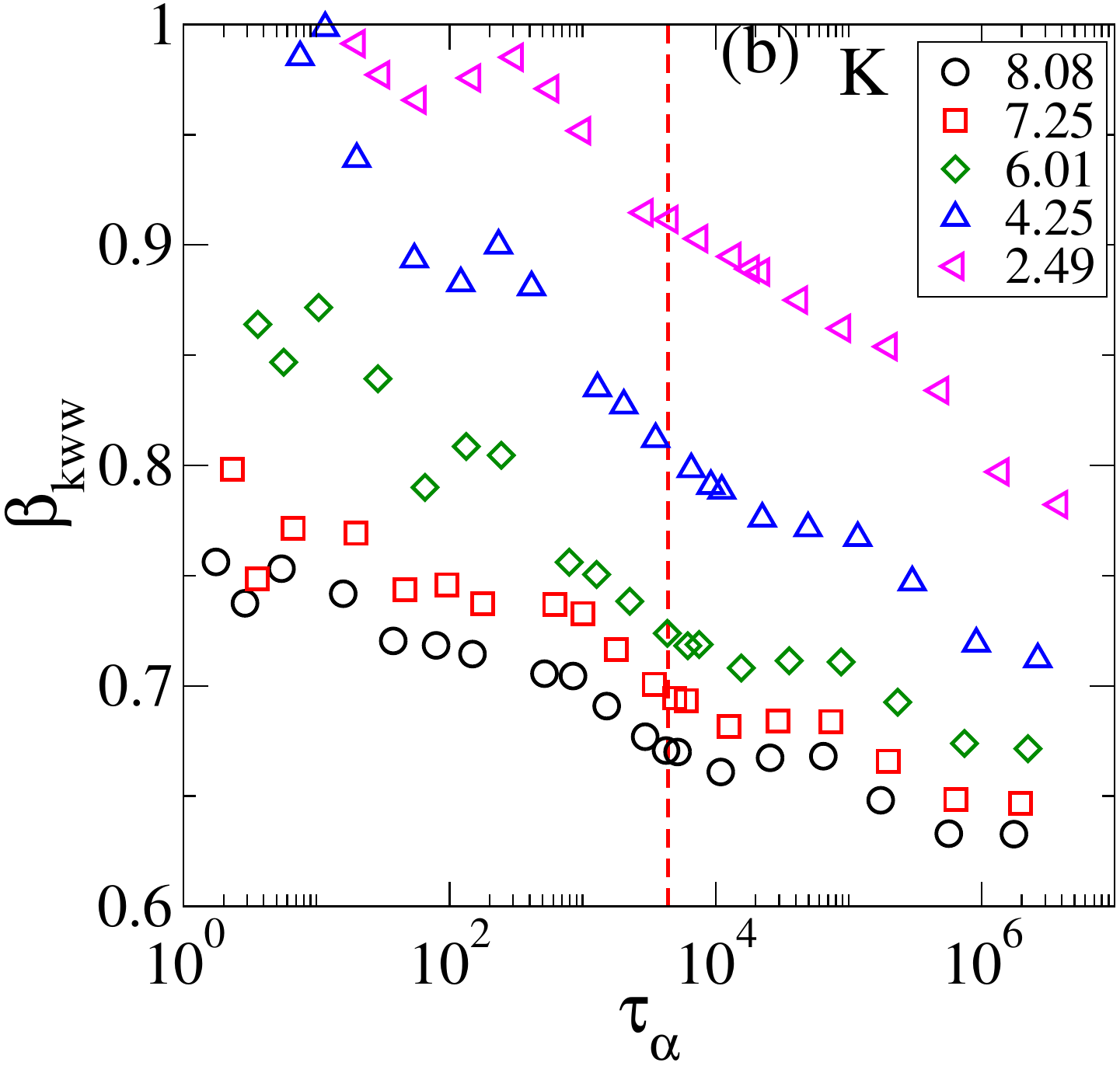}
\includegraphics[scale=0.45]{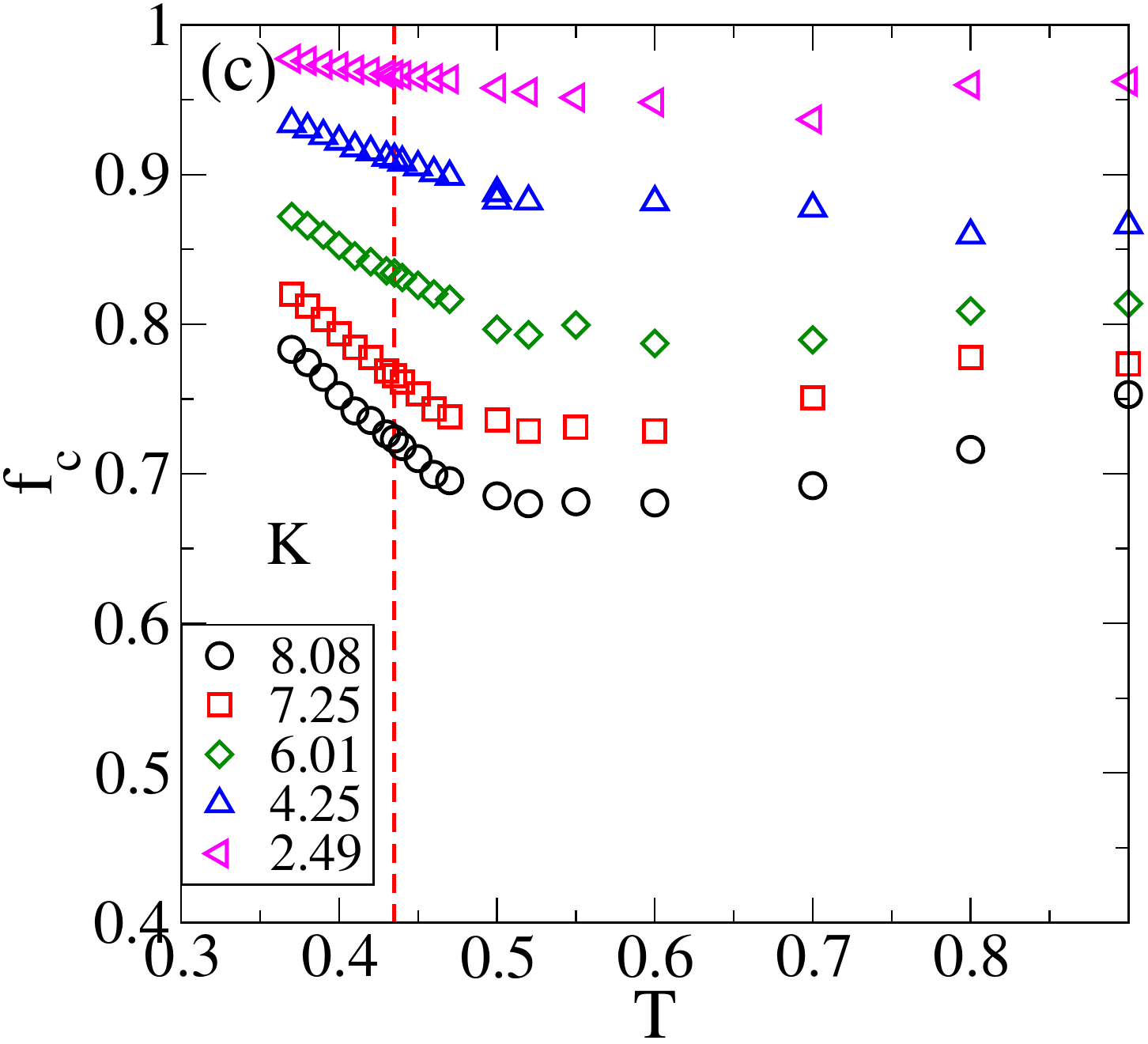}
\caption{The variation of relaxation time $\tau_{\alpha} (k)$, stretching exponent $\beta_{kww}$ and the non-ergodicity parameter ($f_c$), obtained from $F_s(k,t)$, are shown for several $k$ values. 
}
\label{fskt_param}
\end{figure}

The structural relaxation time has also been calculated from the self part of the intermediate scattering function $F{(\textbf{k},t)}$ defined as:\\


\begin{eqnarray}
F_s( \textbf{k}, t) 
       &=& {1 \over N_A} \sum_{i=1}^{N_A} e ^{-i.\textbf{k}.\left(\textbf{r}_i(t_0)-(\textbf{r}_i(t+t_0)\right) } 
\end{eqnarray}

where we calculate $F_s(k,t)$ for the $A$ type of particles, performing an isotropic average over the directions of $\textbf{k}$. 
The structural relaxation time is measured from the $F_s(k,t)$, with $k= |\textbf{k}| = 7.25$, the first peak of the structure factor, unless otherwise noted. $F_s(k,t)$ values we report are obtained by averaging over multiple time origins ($t_0$). The $F_s(k,t)$ curves are shown in Fig. \ref{fsktdat} for several $k$ values, along with fits to the same form as above for $q(t)$. The fit values obtained, of $\tau_{\alpha}(k)$, $\beta_{kww}(k)$ and $f_c(k)$ are shown in Fig. \ref{fskt_param}. 
 
\section{Mean squared displacement}\label{appendixC}

Mean squared displacement (MSD), considering the $A$ particles only, is defined as: 
\begin{equation} 
\triangle r^2(t) = \frac{1}{N_A}\sum_{i = 1}^N|\textbf{r}_i(t+t_o)-\textbf{r}_i(t_o)|^2 
\end{equation} 
where $\textbf{r}_i(t)$ is the position of the $i^{th}$ particle at time $t$. The MSD values are calculated averaging over multiple time origins. The diffusion coefficient $D$ is obtained from the long time behavior of the MSD, from $\triangle r^2(t) = 6 D t$. 

\begin{figure}[h!]
\centering
\includegraphics[scale=0.45]{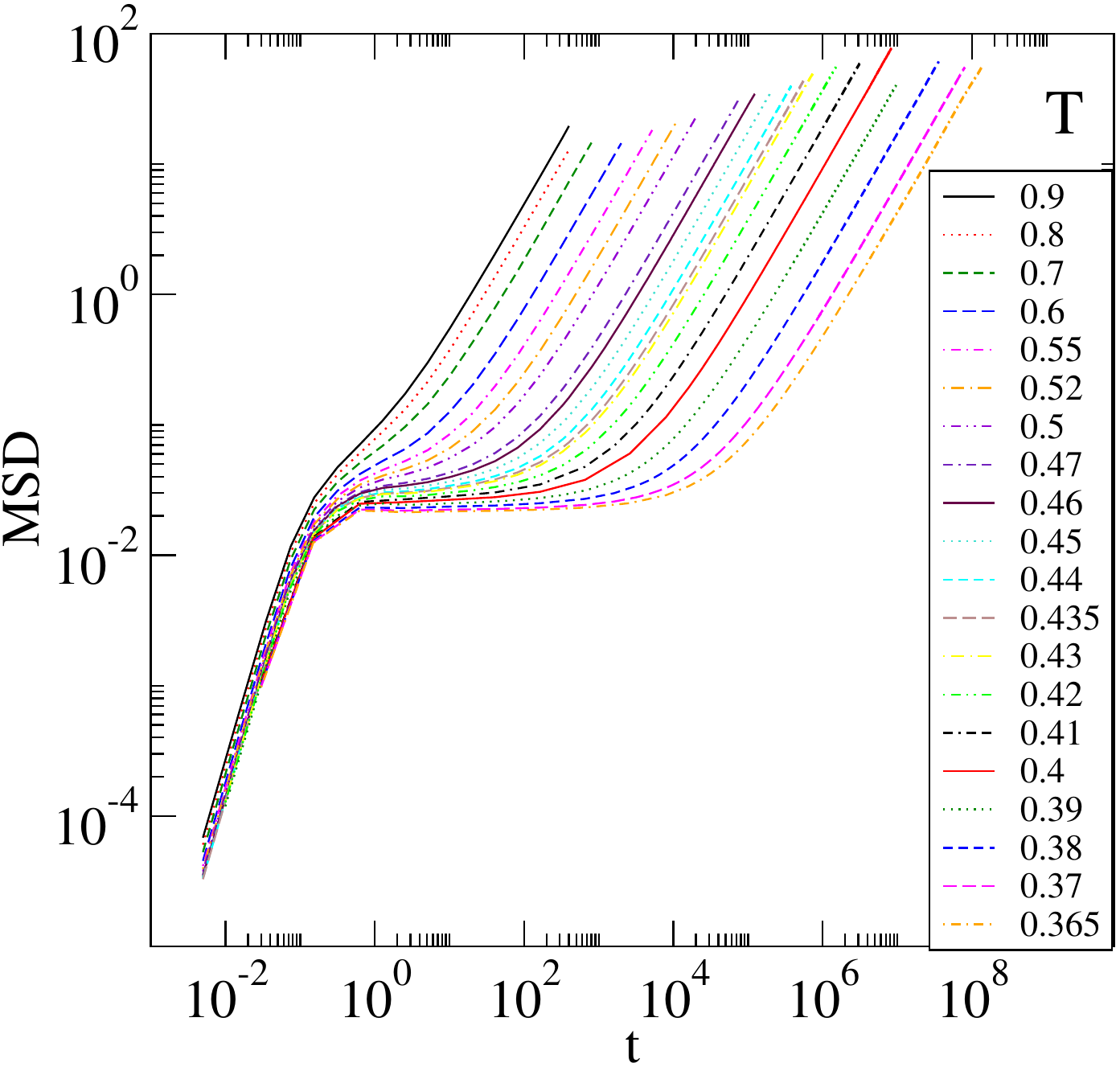}
\caption{Mean squared displacement of the $A$ particles as a function of time for several temperatures.}
\end{figure}


\section{Configurational Entropy} \label{appendixD}


The configurational entropy is obtained by subtracting from the total entropy of the system the vibrational entropy corresponding to the basins of individual glasses.

\begin{equation}
S_{c}=S-S_{vib}
\label{eqsconfs}
\end{equation}

The procedure used to obtain the total entropy, and the vibrational entropy, at a given temperature $T$ and density $\rho$ are described below \cite{ShiladityaThesis2013}.

\subsubsection{Total entropy}

\begin{figure}[h!]
\centering

\includegraphics[scale=0.45]{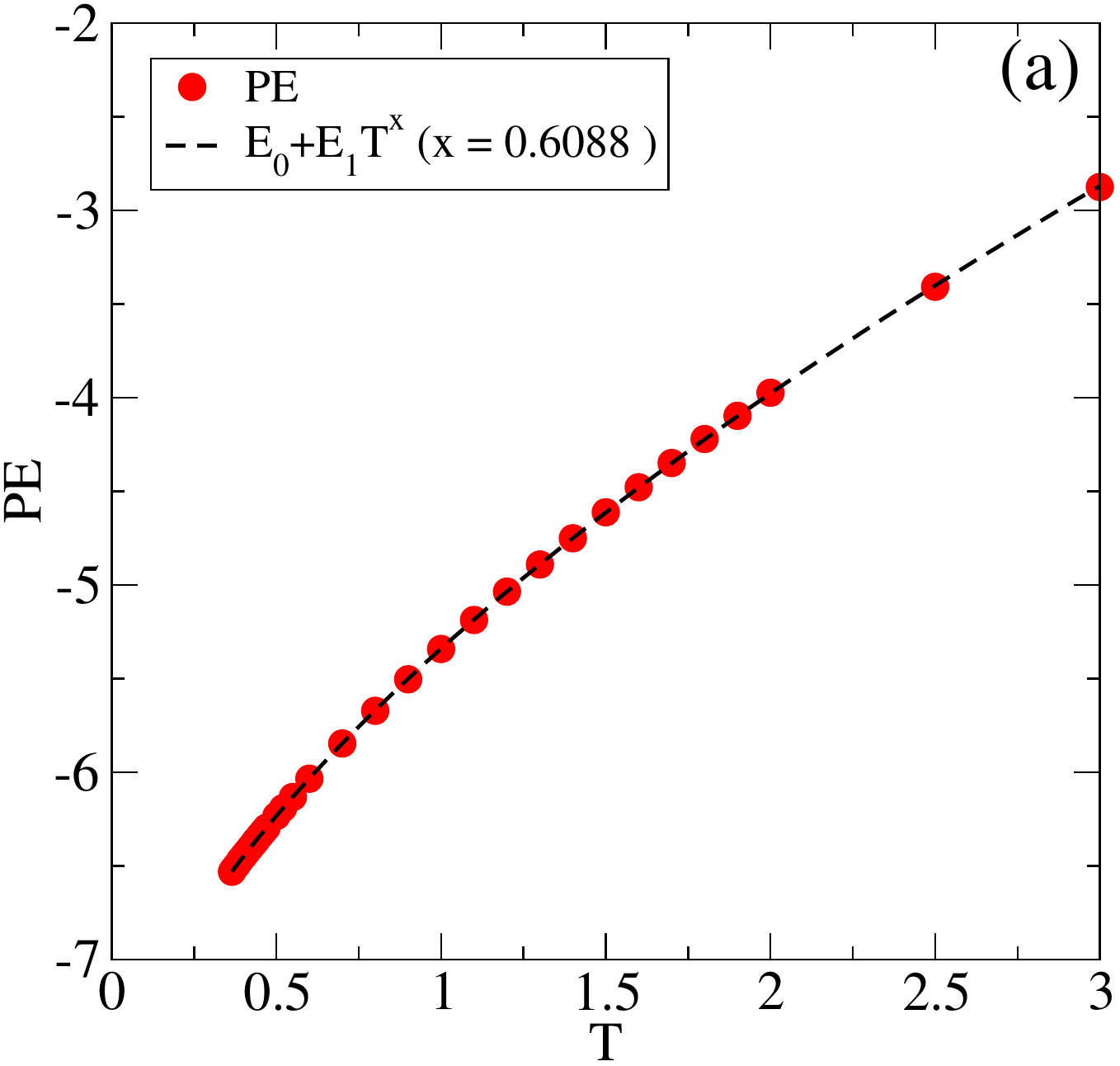}
\includegraphics[scale=0.45]{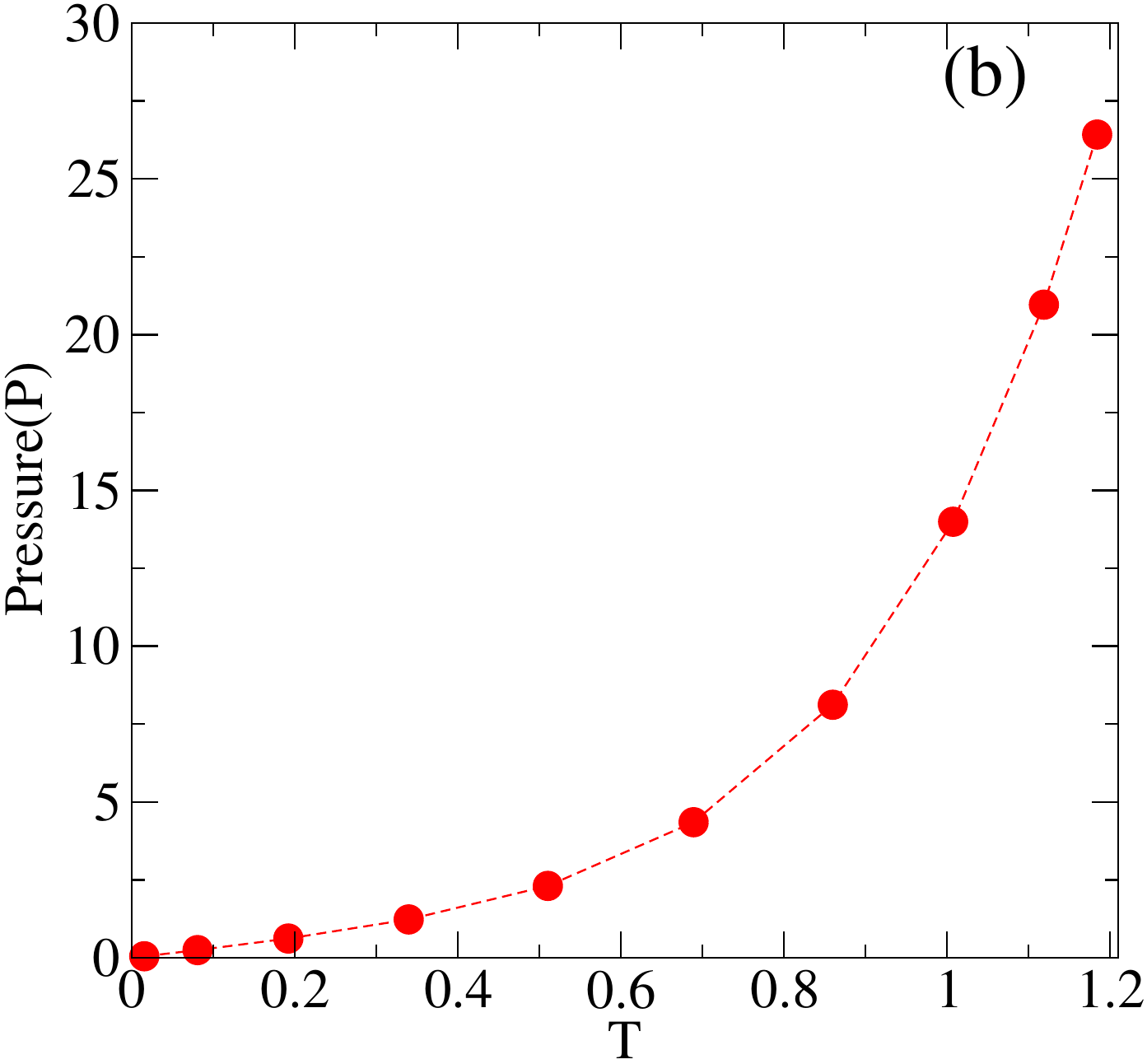}
\caption{ (a) Potential energy at $\rho = 1.2$ as a function of temperature, and (b) pressure as a function of density $\rho$ at reference temperature $T_r = 3.0$. The potential energies are well described by the Rosenfeld-Tarazona scaling of $E = a + b T^{3/5}$, but we use a best fit exponent of $0.6136$ in performing the thermodynamic integration.}
\label{RFscale}
\end{figure}

The total entropy is computed from the Helmholtz free energy $A$, and subtracting the internal energy from it. The Helmholtz free energy is obtained by performing thermodynamic integration numerically from the ideal gas reference state at a high temperature $T_r$, to the desired density, and and a thermodynamic integration at fixed density to the desired low temperature, using the thermodynamic identities: 

\begin{eqnarray}
\left(\frac{\partial A}{\partial V}\right)_{N,T} &=& -P	\\ 
\left(\frac{\partial(A/T)}{\partial(1/T)}\right)_{N,V} &=& U \\
\end{eqnarray}
where $U$ is the internal energy and $P$ is the pressure. We choose a thermodynamic path from zero density to the simulation density of $\rho = 1.2$, at $T_r = 3.0$. Writing the total free energy $A$ as a sum of the ideal and excess parts, we have $A(\rho, T) = A_{id}(\rho,T) + A_{ex} (\rho, T)$, with $A_{id}(\rho,T) = NT(3 ln ~\Lambda + ln~ \rho -1)$,  $\Lambda = {h \over \sqrt{2\pi T} }$. Though not strictly necessary, we use the numerical value of 
the Planck's constant \textmark{$h = 0.1858$} using Argon units, in computing the numbers we report. Thermodynamic integration is performed for the excess free energy, first with respect to density: 

\begin{eqnarray}
A_{ex}(\rho, T_r) - A_{ex}(0, T_r) &=& NT_r\int_0^\rho d\rho \left(\frac{\beta_r P}{\rho^2}-\frac{1}{\rho}\right)\nonumber \\
\end{eqnarray}
where the reference excess free energy is 
\begin{eqnarray}
A_{ex}(0,T_r) = - T_r ln \left(\frac{N!}{N_A!N_B!}\right), 
\end{eqnarray}
and $\beta = \left(k_B T\right)^{-1}$. The pressures employed for the above numerical integration are shown in Fig. \ref{RFscale}.  
The thermodynamic integration to the desired temperature is then performed by integrating the potential energy $E$.
\begin{eqnarray}
A_{ex,T} = T(\beta_rA_{ex}(\rho,T_r)+\int_{\beta_r}^{\beta}d\beta^{'}E(\rho,\beta^{'}))
\end{eqnarray}
The potential energy, shown in Fig. \ref{RFscale}, follows the Rozenfeld-Tarazona scaling $(E = a+bT^{3/5})$ quite well, as noted before \cite{sciortino1999}, but we use the best fit to the numerical data with an exponent of \textmark{$0.6088$} for the numerical integration. The entropy is then obtained from 
\begin{equation}
S  = - \left[ \frac{A(\rho,T)}{T}-\frac{E(\rho,T)}{T}-\frac{3 N}{2} \right].
\end{equation}

\subsubsection{Vibrational entropy (harmonic approximation)}

The vibrational entropy is computed, in the harmonic approximation, by expanding the energy around a local energy minimum of energy $e_{IS}$ as 

\begin{equation}
E = e_{IS} + \frac{1}{2} \sum_{ij\alpha \beta} \frac{  \partial^2 U }{\partial r_{i\alpha} \partial r_{j\beta}} \delta r_{i\alpha} \delta r_{j\beta}, 
\end{equation} 
truncated at the second order.  Diagonalizing the Hessian $H$, with
\begin{equation} 
H_{i\alpha j \beta} = \frac{  \partial^2 E }{\partial r_{i\alpha} \partial r_{j\beta}}
\end{equation}
one obtains $3N - 3$ non-zero eigen values $\lambda_i$ (and $3$ zero eigen values corresponding to global translations), and the corresponding frequencies $\omega_i = \sqrt{\lambda_i}$. The vibrational free energy (with energies measured with respect to the minumum values $e_{IS}$) can be written as 
\begin{equation}
\beta f_{vib} =  \textmark{<}\sum_{i = 1}^{3N-3} \ln \left(\beta \hbar \omega_i\right) - \ln z_0^3 \textmark{>}
\end{equation}
where 
\begin{equation}
z_0 = \left[{1 \over h } \sqrt{{2\pi \over \beta}} V^{1/3}  \right]    
\end{equation}
corresponds to the $3$ zero frequency modes. The vibrational entropy is obtained as 

\begin{equation}
S_{vib} =  - \frac {\partial f_{vib}}{\partial T}.
\end{equation}


\subsubsection{Vibrational entropy (anharmonic correction)}

In the harmonic approximation, the (vibrational) potential energy as a function of temperature can be written as 
\begin{equation}
E_{vib} (T) = \textmark{<}e_{IS}\textmark{>} + {3 N \over 2} k_B T.     
\end{equation}
Thus, if one considers the instantaneous potential energy and subtracts $e_{IS}$ where $e_{IS}$ is the local energy minimum to which the instantaneous configuration maps, the difference should equal ${3 N \over 2} k_B T$. While this is found to be nearly the case for the KA-BMLJ \cite{sastry2000onset}, there is a non-negligible anharmonic component. Thus, if one considers the system to be thermalised within the basin of an  inherent structure, one can write the anharmonic component as 

\begin{equation}
E_{anh} (T) = \textmark{<}E_{vib} (T) - e_{IS}\textmark{>}- {3 N \over 2} k_B T.     
\end{equation}
In order to compute the contribution of this anharmonicity, for inherent structures obtained at a temperature $T_p$, one considers that $E_{anh} (T)$ has a temperature dependence that can be expressed as \cite{Sciortino_2005} 

\begin{equation}
E_{anh} (T) = \sum_{n = 2}^{n_{max}} c_n T^n. 
\label{eanh}
\end{equation}

The derivative of $E_{anh} (T)$ is the anharmonic component of the vibrational specific heat, and therefore one has, with  

\begin{equation}
S_{anh} (T_p) = \int_{T = 0}^{T_p} dT {1 \over T} {\partial E_{anh}(T) \over \partial T},
\end{equation}

\begin{equation}
S_{anh} (T_p) = \sum_{n = 2}^{n_{max}} {n~c_n \over (n-1)} T_p^{n-1}. 
\end{equation} 

\begin{figure}[t]
\centering
\includegraphics[scale=0.45]{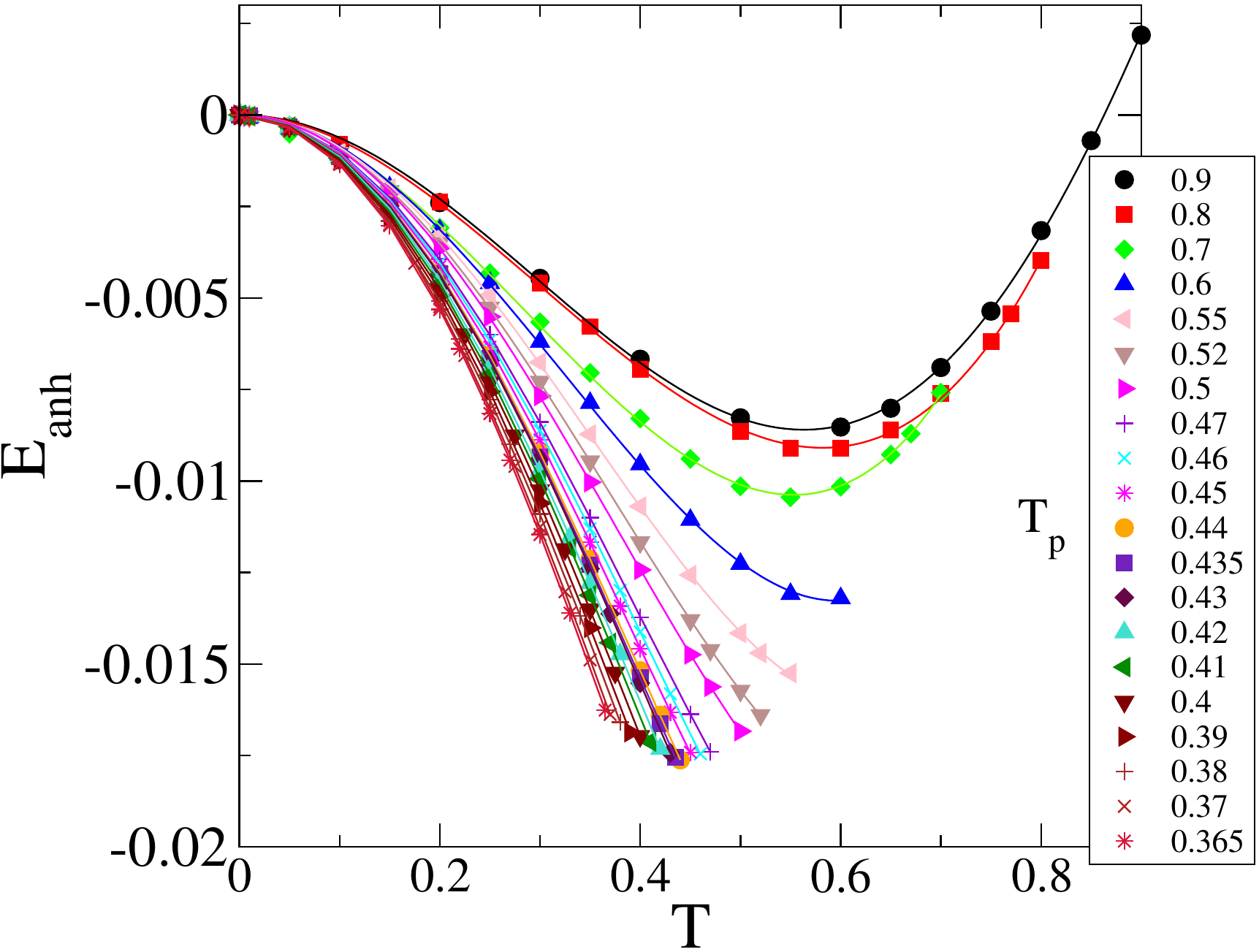}
\caption{Anharmonic corrections to the energy, along with polynomial fits according to Eq. \ref{eanh}.}
\label{eanhfit}
\end{figure}

In order to evaluate this contribution, we consider $1000$ inherent structures at each temperature $T_p$. For each of them, a short simulation (of \textmark{$1000$} integration steps\textmark{, using a time step of $0.005$}) is performed for a range of temperatures $T$ from close to $0$ to $T_p$. The simulation is chosen so that (a) it is longer than the time required for the system to thermalize (as we verify), and (b) not much longer than the caging time, as judged by the plateau of the mean squared displacement, so that a roughly constant energy is obtained in this time window. The energies from \textmark{$150$ to $650$ ($150$ to $500$ for the highest two temperatures) steps are} averaged to obtain estimates of $E_{vib}(T)$, from which $E_{anh}(T)$ is calculated. The resulting data, for each $T_p$ is fitted to the form Eq. \ref{eanh}, \textmark{with $n_{max} = 4$ (except for $T_{p} = 0.9$ for which we use $n_{max} = 5$}, from which $S_{anh}(T_p)$ is computed. \textmark{The fit coeffcients for selected temperatures are shown in Table I below.We show the anharmonic corrections to the energy, along with the fit lines, in Fig. \ref{eanhfit}.}


\begin{table}[]
    \centering
    \begin{tabular}{|c|c|c|c|}
    \hline
        $T_p$ & $c_2$&$c_3$&$c_4$ \\
        \hline
        0.8&-0.0765996& 0.0787235&0.011623\\
        \hline
         0.7&-0.0973854&0.10679&0.0143602\\
\hline
     0.6&-0.0922832&0.0587472&0.056073\\
\hline
0.5&-0.09677&0.00530751&0.10677\\
\hline
0.45&-0.105043&-0.0221009&0.142661\\
\hline
0.4&-0.121862&-0.0144479&0.13501\\
\hline
0.37&-0.134462&-0.00880651&0.131821\\
\hline
    \end{tabular}
    \caption{Fit coefficients $c_n$ in Eq. \ref{eanh} from fits shown in Fig. \ref{eanhfit} for selected temperatures.}
    \label{tab:my_label}
\end{table}

 \bibliographystyle{elsarticle-num} 

 \bibliographystyle{apsrev4-1}
 
 \bibliography{glass}





\end{document}